\documentclass[superscriptaddress,pre,onecolumn,preprintnumbers,amsmath,amssymb]{revtex4-1}
\usepackage{graphicx}
\usepackage{epstopdf, epsfig}
\usepackage{amsmath}
\usepackage{array}
\usepackage{multirow}
\usepackage{psfrag}
\usepackage[font=footnotesize,format=plain,labelsep=period,justification=raggedright]{caption}
\usepackage{subcaption}

\usepackage[utf8]{inputenc}
\usepackage{combelow}
\usepackage{amssymb}
\usepackage{braket}
\usepackage{bm}
\usepackage{color}

\def\bcdot{\cdot}
\def\boldsymbol{\mathbf}
\def\mathsfbi{\mathbf}

\begin{document}

\title{Modelling ternary fluids in contact with elastic membranes}

\author{M. Pepona}
\affiliation{Department of Physics, Durham University, South Road, Durham DH1 3LE, UK}
\author{A.C.M. Shek}
\affiliation{Department of Physics, Durham University, South Road, Durham DH1 3LE, UK}
\author{C. Semprebon}
\affiliation{Smart Materials and Surfaces Laboratory, Department of Mathematics, Physics and Electrical Engineering, Ellison Place, Northumbria University, Newcastle upon Tyne, NE1 8ST, UK}
\author{T. Kr{\"u}ger}
\affiliation{School of Engineering, Institute for Multiscale Thermofluids, The University of Edinburgh, Edinburgh EH9 3FB, Scotland, UK}
\author{H. Kusumaatmaja}
\email{halim.kusumaatmaja@durham.ac.uk}
\affiliation{Department of Physics, Durham University, South Road, Durham DH1 3LE, UK}

\begin{abstract}
We present a thermodynamically consistent model of a ternary fluid interacting with elastic membranes. Following a free-energy modelling approach and taking into account the thermodynamics laws, we derive the equations governing the ternary fluid flow and dynamics of the membranes. We also provide the numerical framework for simulating such fluid-structure interaction problems. It is based on the lattice Boltzmann method, employed for resolving the evolution equations of the ternary fluid in an Eulerian description, coupled to the immersed boundary method, allowing for the membrane equations of motion to be solved in a Lagrangian system. The configuration of an elastic capsule placed at a fluid-fluid interface is considered for validation purposes. Systematic simulations are performed for a detailed comparison with reference numerical results obtained by Surface Evolver, and the Galilean invariance of the proposed model is also proven. The proposed approach is versatile, and a wide range of geometries can be simulated. To demonstrate this, the problem of a capillary bridge formed between two deformable capsules is investigated here.  
\end{abstract}

\maketitle

\section{Introduction}\label{sec:Intro}
Multi-phase/component flows in interaction with deformable, thin structures are encountered in a broad range of applications. One example of such flows, that is abundant in many biological systems, are single/multi-component flows in which soft particles, composed of an elastic shell enclosing one or more substances, are embedded. For instance, red blood cells are comprised of a thin lipid bilayer membrane containing a cytosol with viscosity about five times higher than that of the surrounding plasma \citep{Freund2014}. Likewise, viral capsids are formed of polymer DNA confined by a shell of protein molecules \citep{AlbertsEtAl2002}. In general, capsules accommodating multiple aqueous solutions, such as polyethylene glycol and dextran, serve as model biological cells \citep{ScottLongEtAl2005}. Artificially fabricated elastic capsules \citep{YowRouth2006} are also employed as container and delivery systems in many industrial applications. For example, polymeric multilayer capsules are engineered for drug-delivery purposes \citep{DeCockEtAl2010}, while liquid-core capsules are used in various processes of cosmetic \citep{MartinsEtAl2014} and food \citep{GharsallaouiEtAl2007} industries. Another case where a multicomponent flow interacting with soft particles occurs is in the self-assembly of colloidal aggregates into patchy particles, a phenomenon of particular importance for the successful design of bottom-up materials \citep{GlotzerSolomon2007, PawarKretzschmar2010}. These patchy particles can be formed of compartments, each containing different polymers, bonded together by a solvent. The wetting of liquid droplets surrounded by air/vapour on soft solids is another example of multiphase flows in contact with deformable structures, with relevance in biology, e.g. cell locomotion \citep{LoEtAl2000}, medicine and engineering, e.g. in the identification of cancer cells \citep{DischerEtAl2005}, and in the development of smart coatings \citep{YaoEtAl2013}. Similar elastowetting problems also arise with fibrous materials, such as the coalescence of wet hair \citep{BicoEtAl2004}.

Even though these elastocapillary problems have been extensively studied both analytically and experimentally \citep{IvanovskaEtAl2004, BuenemannLenz2008, HegemannEtAl2018, ScottLongEtAl2008, KusumaatmajaEtAl2009, KusumaatmajaLipowsky2011, MeesterEtAl2016, MarchandEtAl2012, StyleDufresne2012, StyleEtAl2013, BicoEtAl2018, DupratEtAl2011, SinghEtAl2014, SauretEtAl2015}, the accurate computational modelling of such configurations is lagging behind. Few works have dealt with the simulation of elastic liquid-core capsules immersed in another fluid component \citep{RamanujanPozrikidis1998, ZhangEtAl2007, KaouiEtAl2008}. In these works, the ambient and enclosed fluids are assumed to have equal densities and different viscosities, and the fluid-fluid surface tension, if taken into account, takes the form of a force exerted on the membrane, rather than being an inherent quantity of the fluids. There has been an attempt by \citet{LubbersEtAl2014} to numerically solve the equilibrium shape problem of liquid droplets on soft solids. The equilibrium shapes are obtained by minimizing the total elastocapillary energy of the vapour-liquid-substrate system; thus, no fluid dynamics properties of the droplets and surrounding air are available. \citet{ShaoEtAl2013} have developed a numerical technique to study the contact line dynamics of a two-phase fluid in interaction with rigid, solid circular cylinders. Working towards that direction, \citet{LiEtAl2016} have proposed a numerical approach capable of simulating the interaction between a two-phase flow and elastic fibers. In their work, the wetting is controlled by empirical parameters entering the fluid solver through cohesion forces mimicking the role of surface tensions. Recently, \citet{BuenoEtAl2018} have developed a numerical framework for the simulation of binary fluids in contact with nonlinear hyperelastic solids to investigate the wetting of soft substrates and elastic micropillars.
 
The aim of the present work is to analytically derive a thermodynamically consistent model of multicomponent fluids in interaction with elastic membranes, and to provide a versatile numerical framework for the simulation of such fluid-structure interaction problems. We focus here on the case of a ternary fluid where one component is enclosed inside the membranes. The model can, however, be generalised to consider more fluid components contained in/surrounding the membranes. The availability of such computational method will allow us to systematically study a wide range of elastocapillary phenomena intractable to analytical solutions, in order to complement expanding experimental activities in this area.
 
We follow a top-down modelling approach for the ternary fluid using the lattice Boltzmann method \citep{SwiftEtAl1995, SwiftEtAl1996, BriantEtAl2004, BriantYeomans2004, LeeLin2005, MazloomiEtAl2015, SemprebonEtAl2016, WohrwagEtAl2018}; namely, the free energy of the fluid system is initially formulated, including the desired thermodynamics features, such as the immiscibility of the fluid mixture, the surface tensions between the different fluid components, and the energy due to the interaction of the elastic membranes with the confined component. Given the free energy, other thermodynamic properties, e.g. the chemical potentials of each component, and the pressure of the fluid mixture, as well as the macroscopic equations of motion of the multicomponent fluid can be derived. This technique is the contrary of the bottom-up modelling approach, where the macroscopic properties of the fluid arise from the microscopic interactions between the fluid particles. These interactions usually take the form of an interparticle potential \citep{ShanChen1993, ChenEtAl2014}. \citet{KrugerEtAl2017} have discussed in detail the advantages and limitations of each of these modelling approaches. The lattice Boltzmann model proposed by \citet{SemprebonEtAl2016} has been in particular chosen here for resolving the flow in a uniform, Cartesian grid. This is a diffuse interface model, meaning that the fluid-fluid interfaces are spread over several lattices, rather than being tracked explicitly.

The capsules enclosing one fluid component are modelled as infinitely thin membranes, composed of a homogeneous and isotropic material, able to undergo stretching/compression and bending. The response of the membranes to the enclosing and surrounding fluid components is governed by constitutive equations for the strain and bending, whose formulations depend on the nature of the material. Here, we consider the simple case of a linear elastic material, for which the strain energy takes the form of Hooke's law, and the bending energy is given by the squared mean curvature. However, the strain and bending energy formulations can be easily modified to model more realistic materials. For example, the neo-Hookean and Mooney-Rivlin laws are employed for hyperelastic rubber-like materials, while the model of \citet{SkalakEtAl1973} is widely used for red blood cells. 

To resolve the interaction between the ternary fluid and the membranes, we adopt the immersed boundary method (IBM), initially proposed by \citet{Peskin1977}. While the equations of motion of the ternary fluid are solved in an Eulerian description, a Lagrangian coordinates system is used for the solution of the membranes' constitutive laws. The membranes are discretised into points, referred to as Lagrangian markers, connected by linear tension/compression and torsion springs, whose coefficients can be related to the tensile and bending moduli of the desired material. The Lagrangian markers are allowed to move in order to follow the motion/deformation of the membranes' boundary, without having to coincide with the underlying Eulerian fluid lattices. The desired strain and bending conditions of the membranes are then imposed on the Lagrangian markers. By transferring information between the two coordinates systems, the aforementioned Lagrangian conditions take the form of a forcing term, reproducing their effect on the Eulerian flow, in the lattice Boltzmann equation. The information transfer is achieved by interpolation and spreading operations. 

The article is organised as follows. In \S\ref{sec:Model}, we derive a thermodynamically consistent model of a ternary fluid in contact with elastic membranes. The methods employed here for the numerical solution of the ternary fluid and its interacting membranes equations of motion are subsequently presented in \S\ref{subsec:LBM} and \S\ref{subsec:IBM}. The Surface Evolver, an open-source software used for benchmarking purposes, is briefly discussed in \S\ref{subsec:SE}. We then validate our model against Surface Evolver in \S\ref{subsec:LiquidLens}, considering an elastic capsule placed at a fluid-fluid interface as the benchmark configuration. We also prove the Galilean invariance of the equations governing the ternary fluid flow and the membranes' dynamics. To demonstrate the versatility of our model, the problem of a capillary bridge formed between two elastic capsules is afterwards studied in \S\ref{subsec:CapillaryBridges}. It should be noted that although the simulations performed here are restricted to two dimensions, an extension to three-dimensional configurations is straightforward. Finally, the key contributions of the present work and its future perspectives are summarized in \S\ref{sec:Conclusions}. 
 
\section{Mathematical modelling of ternary fluids in interaction with elastic membranes}\label{sec:Model}
In this section, the equations of motion for the ternary fluid interacting with elastic membranes are derived. Following the rationale of \citet{KouSun2018}, we firstly derive an entropy conservation equation for the ternary fluid and the membrane by making use of the first law of thermodynamics and splitting structure of the total entropy into contributions from the system of interest and its surroundings. To further reduce the entropy conservation equation, we then derive a transport equation of the free energy density of the fluid mixture. We subsequently formulate the free and kinetic energies of the elastic membrane. Taking into account the aforementioned and the second law of thermodynamics, we obtain the equations of motion of the ternary fluid and the elastic membrane. We then present briefly the free energy of the ternary fluid, which is based on the one proposed by \citet{SemprebonEtAl2016} but modified accordingly in order to account for the interplay between the fluid and the membrane. By introducing auxiliary fields that allow for an easy numerical implementation of the equations of motion of the ternary fluid, the equivalent formulations of the continuity and Navier-Stokes equations are obtained along with the Cahn-Hilliard equations describing the evolution of these auxiliary fields. 

\subsection{Entropy equation}\label{subsec:Entropy}   
The first law of thermodynamics can be formulated as
\begin{equation}\label{eq:ThermoLaw1}
\frac{d\left(U+E\right)}{dt} = \frac{dW}{dt} + \frac{dQ}{dt},
\end{equation}
where $t$ is time, $U$ and $E$ are respectively the internal and kinetic energies, $W$ is the work done on the ternary fluid system by its surroundings, and $Q$ denotes the amount of heat supplied to the system for its temperature $T$ to be kept constant. The total entropy $S$ can be split into two parts: the entropy of the system $S_{sys}$, and the entropy of its surroundings $S_{sur}$. By definition, the total free energy of the system can be expressed as
\begin{equation}\label{eq:ThermoFE}
F = U-T S_{sys}.
\end{equation}
It is also known that the entropy of the system's surroundings is related to the heat $Q$ by
\begin{equation}\label{eq:Ssur}
dS_{sur} = -\frac{dQ}{T}.
\end{equation}
Taking into account (\ref{eq:ThermoLaw1})--(\ref{eq:Ssur}), it can be shown that
\begin{equation}\label{eq:EntropyRoC}
\frac{dS}{dt} = \frac{dS_{sys}}{dt} + \frac{dS_{sur}}{dt} = -\frac{1}{T}\frac{d\left( F + E \right)}{dt} + \frac{1}{T} \frac{dW}{dt}.
\end{equation}
Here, the system of interest consists of the ternary fluid (denoted by subscript $f$) and the membrane (subscript $m$). Let us first concentrate on the terms related to the ternary fluid.

Given the simulation volume $V$, the free and kinetic energies of the ternary fluid can be defined as
\begin{equation}\label{eq:Densities}
F_f = \int_V f_f\ dV,\ E_f = \frac{1}{2} \int_V \rho |\boldsymbol{u}|^2\ dV,
\end{equation} 
where $f_f$ is the free energy density of the ternary fluid, $\rho$ is the mass density of the fluid mixture, and $\boldsymbol{u}$ denotes the mass-averaged velocity. The fluid free energy density is the sum of two contributions: a free energy density of the bulk fluid $f_b$, and a local free energy density gradient contribution $f_\nabla$ allowing the existence of diffuse fluid-fluid interfaces
\begin{equation}\label{eq:FEdensity}
f_f = f_b + f_\nabla.
\end{equation}
The interfacial free energy density term is defined here as
\begin{equation}\label{eq:FEdensityGrad}
f_\nabla = \frac{1}{2}\sum_{m,n = 1}^N c_{mn} \nabla C_m \bcdot\nabla C_n,
\end{equation}
where $N$ denotes the number of components of the fluid mixture (thus, $N = 3$ here), $c_{mn}$ is the cross influence parameter, and $C_m$ represents the concentration fraction of fluid $m$ $\left( m = 1, 2, 3\right)$. The mass density is given by
\begin{equation}\label{eq:MassDensity}
\rho = \sum_{m = 1}^N C_m M_{w,m},
\end{equation}
where $M_{w,m}$ denotes the weight of component $m$.

By applying the Reynolds transport and Gauss divergence theorems to (\ref{eq:Densities}), we can obtain 
\begin{equation}\label{eq:FERoC}
\frac{dF_f}{dt} = \int_V \frac{\partial f_f}{\partial t}\ dV + \int_V \nabla\bcdot\left( f_f\boldsymbol{u}\right) dV,
\end{equation}
\begin{eqnarray}\label{eq:KERoC}
\frac{dE_f}{dt} & = & \frac{1}{2}\int_V \frac{\partial\left(\rho \boldsymbol{u}\bcdot\boldsymbol{u}\right)}{\partial t}\ dV + \frac{1}{2} \int_V \nabla\bcdot\left(\left(\rho\boldsymbol{u}\bcdot \boldsymbol{u}\right)\boldsymbol{u}\right) dV \nonumber\\
& = & \int_V \rho\boldsymbol{u}\bcdot\left(\frac{\partial \boldsymbol{u}}{\partial t} + \boldsymbol{u}\bcdot\nabla\boldsymbol{u}\right) dV + \frac{1}{2}\int_V \boldsymbol{u}\bcdot\boldsymbol{u}\left( \frac{\partial\rho}{\partial t} + \nabla\bcdot\left(\rho \boldsymbol{u}\right)\right) dV. 
\end{eqnarray}

The mass transfer occurs through the convection and diffusion of each fluid component, when a velocity field is present. The mass conservation law for each component $m$ can then be written as
\begin{equation}\label{eq:MassLawComponent}
\frac{\partial C_m}{\partial t} + \nabla\bcdot\left( C_m\boldsymbol{u}\right) + \nabla\bcdot\boldsymbol{J}_m = 0,
\end{equation}
where $\boldsymbol{J}_m = M_m\nabla\mu_m$ denotes the diffusion flux of component $m$, and $M_m$ represent the corresponding mobility parameters. The chemical potential $\mu_m$ is discussed in \S\ref{subsec:TEFED}. By multiplying (\ref{eq:MassLawComponent}) with $M_{w,m}$, and summing over all fluid components $m$, the mass conservation equation of the fluid mixture can be obtained as
\begin{equation}\label{eq:MassLawFluid}
\frac{\partial\rho}{\partial t} + \nabla\bcdot\left(\rho\boldsymbol{u}\right) + \sum_{m = 1}^N M_{w,m}\nabla\bcdot\boldsymbol{J}_m = 0.
\end{equation}
By substituting (\ref{eq:MassLawFluid}) into (\ref{eq:KERoC}), the latter can be rewritten as
\begin{eqnarray}\label{eq:KERoCv2}
\frac{dE_f}{dt} & = & \int_V \rho\boldsymbol{u}\bcdot\frac{d \boldsymbol{u}}{dt}\ dV -\frac{1}{2}\int_V \sum_{m = 1}^N M_{w,m}\left(\nabla\bcdot\boldsymbol{J}_m\right)\left(\boldsymbol{u}\bcdot
\boldsymbol{u}\right) dV\nonumber\\
& = & \int_V \boldsymbol{u}\bcdot\left(\rho\frac{d\boldsymbol{u}}{dt} + \sum_{m = 1}^N M_{w,m}\boldsymbol{J}_m\bcdot\nabla\boldsymbol{u}\right) dV \nonumber\\
&& -\frac{1}{2}\int_V \sum_{m = 1}^N M_{w,m}\nabla\bcdot\left( \left(\boldsymbol{u}\bcdot\boldsymbol{u}\right)\boldsymbol{J}_m\right) dV,
\end{eqnarray}
where $\frac{d\boldsymbol{u}}{dt}$ is the total derivative of $\boldsymbol{u}$, defined as $\frac{\partial\boldsymbol{u}}{\partial t} + \boldsymbol{u}\bcdot \nabla\boldsymbol{u}$.

The rate of change of the work done by the force $\boldsymbol{F}_{sur}$ is given by
\begin{equation}\label{eq:WRoC}
\frac{dW}{dt} = \int_{\partial V} \left(\boldsymbol{F}_{sur}\bcdot \boldsymbol{u}\right) ds,
\end{equation}
where the integration takes place over the volume's surface boundary $\partial V$. The force $\boldsymbol{F}_{sur}$ is related to the Cauchy stress tensor $\boldsymbol{\sigma}$ of the ternary fluid by $\boldsymbol{F}_ {sur} = -\boldsymbol{\sigma}\bcdot\boldsymbol{n}$, where $\boldsymbol{n}$ denotes the outward unit normal vector of $V$. Equation (\ref{eq:WRoC}) can thus take the form
\begin{eqnarray}\label{eq:WRoCv2}
\frac{dW}{dt} & = & -\int_{\partial V} \left(\left(\boldsymbol{\sigma} \bcdot\boldsymbol{n}\right)\bcdot\boldsymbol{u}\right)ds\nonumber\\
& = & -\int_V \left(\boldsymbol{\sigma}^T\boldsymbol{:}\nabla \boldsymbol{u} + \boldsymbol{u}\bcdot\left(\nabla\bcdot\boldsymbol{\sigma} \right)\right) dV.
\end{eqnarray}

By making use of (\ref{eq:FERoC}), (\ref{eq:KERoCv2}) and (\ref{eq:WRoCv2}), the terms on the right-hand side of (\ref{eq:EntropyRoC}) related to the ternary fluid can be collected as follows
\begin{eqnarray}\label{eq:EntropyLaw}
-\frac{d\left( F_f + E_f \right)}{dt} + \frac{dW}{dt} & = & \int_V\left[ -\frac{\partial f_f}{\partial t} - \nabla\bcdot\left( f_f \boldsymbol{u}\right) \right.\nonumber\\
&& \left. + \frac{1}{2}\sum_{m = 1}^N M_{w,m} \nabla\bcdot\left(\left(\boldsymbol{u}\bcdot\boldsymbol{u}\right)\boldsymbol{J}_ m\right) -\boldsymbol{\sigma}^T\boldsymbol{:}\nabla \boldsymbol{u} \right.\\
&& \left. -\boldsymbol{u}\bcdot\left(\rho\frac{d\boldsymbol{u}}{dt} + \sum_{m = 1}^N M_{w,m}\boldsymbol{J}_m\bcdot\nabla\boldsymbol{u} + \nabla\bcdot \boldsymbol{\sigma}\right)\right] dV.\nonumber
\end{eqnarray} 

\subsection{Transport equation of the fluid free energy density}\label{subsec:TEFED}
The pressure $p$ of the fluid mixture can be defined as
\begin{equation}\label{eq:Pressure}
p = \sum_{m = 1}^N C_m\mu_m - f_f,
\end{equation}
where $\mu_m$ is the chemical potential of component $m$, given by
\begin{equation}\label{eq:ChemPot}
\mu_m = \frac{\delta f_f}{\delta C_m} = \mu_m^b - \underbrace{\sum_{n = 1}^N \nabla\bcdot\left( c_{mn}\nabla C_n\right)}_{\mu_m^\nabla},
\end{equation}
where $\mu_m^b = \frac{\delta f_b}{\delta C_m}$ and $\mu_m^\nabla = \frac{\delta f_\nabla}{\delta C_m}$ stand for the bulk and interfacial chemical potentials of component $m$. Due to (\ref{eq:ChemPot}), equation (\ref{eq:Pressure}) can be rewritten as
\begin{equation}\label{eq:PressureV2}
p = \underbrace{\sum_{m = 1}^N C_m\mu_m^b - f_b}_{p_b} - \sum_{m,n = 1}^N C_m \nabla\bcdot\left( c_{mn}\nabla C_n\right) - \frac{1}{2} \sum_{m,n = 1}^N c_{mn} \nabla C_m\bcdot\nabla C_n.
\end{equation}
The bulk pressure $p_b$ can be further divided into two parts: $p_{b,f}$ accounting for the contribution of the bulk free energy density owing to the chosen free-energy functional $f_{b,f}$ of the fluid mixture, and $p_{b,c}$ taking into account the contribution of the bulk free energy density due to a fluid-membrane coupling-energy functional $f_{b,c}$
\begin{equation}\label{eq:BulkPres}
p_b = \underbrace{\sum_{m = 1}^N C_m \mu_{m}^{b,f}-f_{b,f}}_{p_{b,f}} + \underbrace{C_3\frac{\delta f_{b,c}\left( C_3,\mathcal{I}\right)}{\delta C_3} - f_{b,c}}_{p_{b,c}},
\end{equation}
where $f_b = f_{b,f}+f_{b,c}$ and $\mu_m^{b,f} = \frac{\delta f_{b,f}\left( C_m\right)}{\delta C_m}$. The coupling-energy functional $f_{b,c}$ depends only on the component enclosed in the membrane, which is assumed to be fluid $3$ in this work, and an interfacial profile $\mathcal{I}$ across the elastic membrane for the ternary fluid. The latter will be discussed in more detail in \S\ref{subsec:FluidFE}. The gradient of the bulk pressure can be found as
\begin{eqnarray}\label{eq:GradBulkPres}
\nabla p_b & = & \nabla\left(\sum_{m = 1}^N C_m\mu_m^b -f_b\right)\nonumber\\
& = & \sum_{m = 1}^N \left( \mu_m^b\nabla C_m + C_m\nabla\mu_m^b\right) - \sum_{m = 1}^N \mu_m^b\nabla C_m - \frac{\delta f_b}{\delta\mathcal{I}}\nabla\mathcal{I} \nonumber\\
& = & \sum_{m = 1}^N C_m\nabla\mu_m^b - \frac{\delta f_{b,c}}{\delta\mathcal{I}} \nabla\mathcal{I}.
\end{eqnarray}
It can also be shown that
\begin{equation}\label{eq:GradFbI}
\nabla p_{b,c} - C_3\nabla\left(\frac{\delta f_{b,c}}{\delta C_3}\right) = \nabla\left( C_3\frac{\delta f_{b,c}}{\delta C_3} - f_{b,c}\right) - C_3\nabla\left(\frac{\delta f_{b,c}}{\delta C_3}\right) = -\frac{\delta f_{b,c}}{\delta\mathcal{I}}\nabla\mathcal{I}.
\end{equation}

Taking into account (\ref{eq:FEdensityGrad}), (\ref{eq:MassLawComponent}), (\ref{eq:GradBulkPres}) and (\ref{eq:GradFbI}), the time derivative of the bulk $f_b$ and interfacial $f_\nabla$ free energy densities can be deduced as
\begin{eqnarray}\label{eq:DtBulkFED}
\frac{\partial f_b}{\partial t} & = &\sum_{m = 1}^N \mu_m^b\frac{\partial C_m}{\partial t} + \frac{\delta f_b}{\delta\mathcal{I}}\frac{\partial\mathcal{I}}{\partial t} = -\sum_{m = 1}^N \mu_m^b\left(\nabla\bcdot\left( C_m\boldsymbol{u}\right) + \nabla\bcdot\boldsymbol{J}_m\right) + \frac{\delta f_{b,c}}{\delta\mathcal{I}}\frac{\partial\mathcal{I}}{\partial t}\nonumber\\
& = & -\nabla\bcdot\sum_{m = 1}^N C_m\mu_m^b\boldsymbol{u} + \sum_{m = 1}^N C_m\boldsymbol{u}\bcdot\nabla\mu_m^b -\sum_{m = 1}^N\mu_m^b\nabla\bcdot \boldsymbol{J}_m + \frac{\delta f_{b,c}}{\delta\mathcal{I}} \frac{\partial\mathcal{I}}{\partial t} \nonumber\\
& = & -\nabla\bcdot\sum_{m = 1}^N\left( C_m\mu_m^b\boldsymbol{u} - f_b \boldsymbol{u}\right) - \nabla\bcdot \left( f_b\boldsymbol{u}\right) + \sum_{m = 1}^N C_m\boldsymbol{u}\bcdot\nabla\mu_m^b -\sum_{m = 1}^N\mu_m^b\nabla\bcdot \boldsymbol{J}_m \nonumber\\
&& + \frac{\delta f_{b,c}}{\delta\mathcal{I}} \frac{\partial\mathcal{I}}{\partial t}\nonumber\\
& = & -\nabla\bcdot\left( p_b\boldsymbol{u}\right) - \nabla\bcdot \left( f_b\boldsymbol{u}\right) + \sum_{m = 1}^N C_m\boldsymbol{u}\bcdot\nabla\mu_m^b -\sum_{m = 1}^N\mu_m^b\nabla\bcdot \boldsymbol{J}_m + \frac{\delta f_{b,c}}{\delta\mathcal{I}} \frac{\partial\mathcal{I}}{\partial t}\nonumber\\
& = & - p_b\nabla\bcdot\boldsymbol{u} - \boldsymbol{u}\bcdot\left( \nabla p_{b,c} - C_3\nabla\left(\frac{\delta f_{b,c}}{\delta C_3}\right)\right) - \nabla\bcdot \left( f_b\boldsymbol{u}\right) -\sum_{m = 1}^N\mu_m^b\nabla\bcdot \boldsymbol{J}_m \nonumber\\
&& + \frac{\delta f_{b,c}}{\delta\mathcal{I}} \frac{\partial\mathcal{I}}{\partial t}
\end{eqnarray}
and
\begin{eqnarray}\label{eq:DtInterFED}
\frac{\partial f_\nabla}{\partial t} & = & \frac{1}{2}\frac{\partial\left( \sum_{m,n = 1}^N c_{mn}\nabla C_m\bcdot\nabla C_n\right)}{\partial t}  = \sum_{m,n = 1}^N c_{mn}\nabla C_m\bcdot\nabla\left(\frac{\partial C_n}{\partial t}\right)\nonumber\\
& = & -\sum_{m,n = 1}^N c_{mn}\nabla C_m\bcdot\nabla\left( \nabla\bcdot\left( C_n\boldsymbol{u}\right) + \nabla\bcdot\boldsymbol{J}_n\right)\nonumber\\
& = & -\sum_{m,n = 1}^N\nabla\bcdot\left(\left(\nabla\bcdot\left( C_n\boldsymbol{u}\right)\right) c_{mn}\nabla C_m\right) + \sum_{m,n = 1}^N\left( \boldsymbol{u}\bcdot\nabla C_n\right)\nabla\bcdot\left( c_{mn}\nabla C_m\right) \nonumber\\
&& + \sum_{m,n = 1}^N C_n\left(\nabla\bcdot\boldsymbol{u}\right)\nabla\bcdot \left( c_{mn}\nabla C_m\right) - \sum_{m,n = 1}^N\nabla\bcdot\left(\left( \nabla\bcdot\boldsymbol{J}_n\right) c_{mn}\nabla C_m \right)\nonumber\\
&& + \sum_{m,n = 1}^N\left(\nabla\bcdot\boldsymbol{J}_n\right)\nabla\bcdot\left( c_{mn}\nabla C_m\right).
\end{eqnarray}
The divergence of the interfacial free energy density in the presence of a velocity field is given by
\begin{eqnarray}\label{eq:DivInterFED}
\hspace*{-20pt}
\nabla\bcdot\left( f_\nabla\boldsymbol{u}\right) & = &  \frac{1}{2} \nabla\bcdot \left(\sum_{m,n = 1}^N c_{mn}\nabla C_m\bcdot\nabla C_n\boldsymbol{u} \right)\\
& = & \frac{1}{2}\left(\sum_{m,n = 1}^N c_{mn}\nabla C_m\bcdot\nabla C_n \right)\nabla\bcdot\boldsymbol{u} + \frac{1}{2}\boldsymbol{u}\bcdot\nabla\left( \sum_{m,n = 1}^N c_{mn}\nabla C_m\bcdot\nabla C_n\right).\nonumber
\end{eqnarray}

Making use of (\ref{eq:FEdensity}), (\ref{eq:ChemPot}), (\ref{eq:PressureV2}) and (\ref{eq:DtBulkFED})--(\ref{eq:DivInterFED}), the transport equation of the free energy density $f_f$ of the ternary fluid reads
\begin{eqnarray}\label{eq:TEFED}
\frac{\partial f_f}{\partial t} + \nabla\bcdot\left( f_f\boldsymbol{u}\right) & = & \frac{\partial f_b}{\partial t} + \nabla\bcdot\left( f_b\boldsymbol{u} \right) + \frac{\partial f_\nabla}{\partial t} + \nabla\bcdot\left( f_\nabla \boldsymbol{u}\right)\nonumber\\
& = & -p\nabla\bcdot\boldsymbol{u} - \sum_{m = 1}^N \mu_m\nabla\bcdot \boldsymbol{J}_m - \sum_{m,n = 1}^N \nabla\bcdot\left(\left(\nabla\bcdot \left(C_n\boldsymbol{u}\right)\right) c_{mn}\nabla C_m\right)\nonumber\\
&& + \sum_{m,n = 1}^N \left(\boldsymbol{u}\bcdot\nabla C_m\right)\nabla\bcdot \left( c_{mn}\nabla C_n\right) - \sum_{m,n = 1}^N\nabla\bcdot\left(\left( \nabla\bcdot\boldsymbol{J}_n\right) c_{mn}\nabla C_m\right)\nonumber\\
&& + \frac{1}{2}\boldsymbol{u}\bcdot\sum_{m,n = 1}^N\nabla\left( c_{mn}\nabla C_m\bcdot\nabla C_n\right) -\boldsymbol{u}\bcdot\left(\nabla p_{b,c} - C_3\nabla \left(\frac{\delta f_{b,c}}{\delta C_3}\right)\right) \nonumber\\
&& + \frac{\delta f_{b,c}}{\delta\mathcal{I}}\frac{\partial\mathcal{I}}{\partial t}.
\end{eqnarray}
Taking into account that the following identity holds
$$\sum_{m,n = 1}^N \left(\nabla C_m\right)\nabla\bcdot\left( c_{mn}\nabla C_n\right) + \frac{1}{2}\sum_{m,n = 1}^N\nabla\left( c_{mn}\nabla C_m\bcdot \nabla C_n\right) = \sum_{m,n = 1}^N \nabla\bcdot\left( c_{mn}\nabla C_m\otimes\nabla C_n\right),$$
equation (\ref{eq:TEFED}) becomes
\begin{eqnarray}\label{eq:TEFEDv2}
\frac{\partial f_f}{\partial t} + \nabla\bcdot\left( f_f\boldsymbol{u}\right) & = & -p\nabla\bcdot\boldsymbol{u} - \sum_{m,n = 1}^N\nabla\bcdot\left(\left( \nabla\bcdot\left( C_n\boldsymbol{u} \right)\right) c_{mn}\nabla C_m\right) - \sum_{m = 1}^N\nabla\bcdot\left(\mu_m \boldsymbol{J}_m\right) \nonumber\\
&&  + \sum_{m = 1}^N\boldsymbol{J}_m\bcdot\nabla\mu_m + \nabla\bcdot\left(\sum_{m,n = 1}^N c_{mn}\left(\nabla C_m\otimes\nabla C_n\right)\bcdot\boldsymbol{u} \right)\nonumber\\
&& - \left(\sum_{m,n = 1}^N c_{mn}\left(\nabla C_m\otimes\nabla C_n\right)\right) \boldsymbol{:}\nabla\boldsymbol{u} - \sum_{m,n = 1}^N\nabla\bcdot\left( \left(
\nabla\bcdot\boldsymbol{J}_n\right) c_{mn}\nabla C_m\right) \nonumber\\
&& -\boldsymbol{u}\bcdot\left(\nabla p_{b,c} - C_3\nabla \left(\frac{\delta f_{b,c}}{\delta C_3}\right)\right)+ \frac{\delta f_{b,c}}{\delta\mathcal{I}} \frac{\partial\mathcal{I}}{\partial t}. 
\end{eqnarray}
By substituting (\ref{eq:TEFEDv2}) into (\ref{eq:EntropyLaw}), the latter is reformulated as
\begin{eqnarray}\label{eq:TEFEDv3}
-\frac{d\left( F_f + E_f\right)}{dt} + \frac{dW}{dt} & = & \int_V\left[\left( p\mathsfbi{I} + \sum_{m,n = 1}^N c_{mn}\left( \nabla C_m\otimes\nabla C_n\right) \right)\boldsymbol{:}\nabla\boldsymbol{u} \right.\nonumber\\ 
&& + \sum_{m = 1}^N \nabla\bcdot\left(\mu_m\boldsymbol{J}_m\right) - \sum_{m = 1}^N \boldsymbol{ J}_m\bcdot\nabla\mu_m \nonumber\\
&& + \sum_{m,n = 1}^N \nabla \bcdot\left(\left(\nabla\bcdot\left( C_n\boldsymbol{u}\right)\right) c_{mn}\nabla C_m\right) \nonumber\\
&& - \nabla\bcdot\left(\sum_{m,n = 1}^N c_{mn}\left(\nabla C_m\otimes\nabla C_n \right)\bcdot\boldsymbol{u}\right)\nonumber\\
&& + \sum_{m,n = 1}^N \nabla\bcdot\left(\left( \nabla\bcdot\boldsymbol{J}_n\right) c_{mn}\nabla C_m\right)\\
&& +\boldsymbol{u}\bcdot\left(\nabla p_{b,c} - C_3\nabla \left(\frac{\delta f_{b,c}}{\delta C_3}\right)\right) - \frac{\delta f_{b,c}}{\delta\mathcal{I}} \frac{\partial\mathcal{I}}{\partial t}\nonumber\\
&& +\frac{1}{2}\sum_{m = 1}^N M_{w,m}\nabla\bcdot\left(\left(\boldsymbol{u} \bcdot\boldsymbol{u}\right) \boldsymbol{J}_m\right) - \boldsymbol{\sigma}^T \boldsymbol{:}\nabla\boldsymbol{u} \nonumber\\
&& \left. - \boldsymbol{u}\bcdot\left(\rho\frac{d\boldsymbol{u}}{d t} + \sum_{m = 1}^N M_{w,m}\boldsymbol{J}_m\bcdot\nabla\boldsymbol{u}+\nabla\bcdot\boldsymbol{\sigma} \right)\right] dV,\nonumber
\end{eqnarray}
where $\mathsfbi{I}$ is the second-order identity tensor. 

\subsection{Free and kinetic energies of the elastic membrane}\label{subsec:EoMMembrane}
Let us focus now on the terms of (\ref{eq:EntropyRoC}) related to the membrane. The free energy $F_m$ of the membrane is the sum of two contributions:  strain energy $\mathcal{E}_s$ and bending energy $\mathcal{E}_b$,
\begin{equation}\label{eq:FEMembrane}
F_m = \mathcal{E}_s + \mathcal{E}_b,
\end{equation}
where
\begin{equation}\label{eq:StretchingE}
\mathcal{E}_s = \int_S \frac{\kappa_s}{2}\left( \left\vert \frac{\partial\boldsymbol{X}}{\partial\boldsymbol{s}}\right\vert - 1\right)^2 d\boldsymbol{s},
\end{equation}
\begin{equation}\label{eq:BendingE}
\mathcal{E}_b = \int_S\frac{\kappa_b}{4} \kappa^2 d\boldsymbol{s}^\prime.
\end{equation}
The integration takes place over the surface $S$ of the elastic membrane, whose position in Eulerian coordinates at time $t$ is described by $\boldsymbol{X} = \boldsymbol{X}\left(\boldsymbol{s}, t\right)$, where $\boldsymbol{s}$ denotes its Lagrangian coordinates. The parameters $\kappa_s$ and $\kappa_b$ are, respectively, the Young's and bending moduli. The variable $\kappa = \partial\theta/\partial\boldsymbol{s}^\prime$, where $\theta = \theta\left(\boldsymbol{X}\right)$ is the tangential angle of an arc of length $\boldsymbol{s}^\prime = \boldsymbol{s}^\prime\left(\boldsymbol{X}\right)$, denotes the curvature of the surface $S$. The kinetic energy $E_m$ of the membrane is given by 
\begin{equation}\label{eq:KineticE}
E_m = \frac{1}{2}\int_S m \left\vert\frac{d\boldsymbol{X}}{d t}\right\vert^2 d\boldsymbol{s},
\end{equation}
where $m$ is the mass of the membrane, here set to be equal to $1$.

\subsection{Equations of motion}\label{subsec:EoMFluid}
By combining equations (\ref{eq:TEFEDv3}) and (\ref{eq:FEMembrane})--(\ref{eq:KineticE}), the entropy equation (\ref{eq:EntropyRoC}) can be rewritten as
\begin{eqnarray}\label{eq:TDToralEntropy}
T\frac{dS}{dt} & = & -\frac{d\left( F_f + E_f\right)}{dt} - \frac{d\left( F_m + E_m\right) }{dt} + \frac{dW}{dt} \nonumber\\
& = & -\int_V \left(\boldsymbol{\sigma}^T -p\mathsfbi{I}-\sum_{m,n = 1}^N c_{mn}\left(\nabla C_m\otimes\nabla C_n \right) \right)\boldsymbol{:}\nabla\boldsymbol{u}\ dV - \int_V \sum_{m = 1}^N\boldsymbol{J}_m\bcdot\nabla\mu_m\ dV \nonumber\\
&& - \int_V \boldsymbol{u}\bcdot\left(\rho\frac{d\boldsymbol{u}}{dt} + \sum _{m = 1}^N M_{w,m}\boldsymbol{J}_m\bcdot\nabla\boldsymbol{u} + \nabla\bcdot\boldsymbol {\sigma} - \nabla p_{b,c} + C_3\nabla \left(\frac{\delta f_{b,c}}{\delta C_3}\right) \right) dV\nonumber\\
&& -\int_V \frac{\delta f_{b,c}}{\delta\mathcal{I}}\frac{\partial\mathcal{I}}{\partial t}\ dV 
- \frac{d\mathcal{E}_s}{dt} 
- \frac{d\mathcal{E}_b}{dt} 
- \int_S m \frac{d\boldsymbol{X}}{dt}\frac{d^2\boldsymbol{X}}{dt^2}\ d\boldsymbol{s}.
\end{eqnarray}
Here, we have used the natural boundary conditions
\begin{equation}\label{eq:NaturalBCs}
\boldsymbol{u}\bcdot\boldsymbol{n}_{\partial V} = 0,\ \boldsymbol{J}_m\bcdot \boldsymbol{n}_{\partial V} = 0,\ \nabla C_m\bcdot\boldsymbol{n}_{\partial V} = 0,
\end{equation}
where $\boldsymbol{n}_{\partial V}$ denotes the outward unit vector normal to the surface boundary $\partial V$ when integrating over the volume $\Omega$ of the fluid mixture. 

The stress tensor $\boldsymbol{\sigma}$ of the ternary fluid can be considered as the sum of two contributions: a reversible stress tensor $\boldsymbol{\sigma}_ {rev}$, and an irreversible one $\boldsymbol{\sigma}_{irrev}$
\begin{equation}\label{eq:StressTensor}
\boldsymbol{\sigma} = \boldsymbol{\sigma}_{rev} + \boldsymbol{\sigma}_{irrev}.
\end{equation}
In an ideal reversible process, there are no effects of viscosity and friction, the diffusion fluxes $\boldsymbol{J}_m$ are equal to $0$, and the entropy is conserved. Taking these facts into consideration, equation (\ref{eq:TDToralEntropy}) results in
\begin{equation}\label{eq:RevST}
\boldsymbol{\sigma}_{rev} = p\mathsfbi{I} + \sum_{m,n = 1}^N c_{mn}\left(\nabla C_m\otimes\nabla C_n\right).
\end{equation}
For an incompressible viscous Newtonian fluid, the irreversible stress tensor can be expressed as
\begin{equation}\label{eq:IrrevST}
\boldsymbol{\sigma}_{irrev} = -\eta\left(\nabla\boldsymbol{u} + \nabla\boldsymbol {u}^T\right),
\end{equation}
where $\eta$ is the dynamic viscosity.

According to the second law of thermodynamics, the total entropy cannot decrease over time. This, in combination with the non-negative nature of the third term on the right-hand side of (\ref{eq:TDToralEntropy}), implies that
\begin{equation}\label{eq:MEv1}
\rho\frac{d\boldsymbol{u}}{dt} + \sum_{m = 1}^N M_{w,m}\boldsymbol{J}_m\bcdot \nabla\boldsymbol{u} + \nabla\bcdot\boldsymbol{\sigma} - \nabla p_{b,c} + C_3 \nabla\left(\frac{\delta f_{b,c}}{\delta C_3}\right) = 0.
\end{equation}
In the present work, we consider that the weight of each fluid component is equal to $1$, namely $M_{w,m} = 1,\ m = 1,2,3$, and $\sum_{m = 1}^N M_{w,m}\boldsymbol {J}_m = 0$. As such, the continuity (\ref{eq:MassLawFluid}) and the Navier-Stokes (\ref{eq:MEv1}) equations become
\begin{equation}\label{eq:ContEq}
\frac{\partial\rho}{\partial t} + \nabla\bcdot\left(\rho\boldsymbol{u}\right) = 0, 
\end{equation}
and
\begin{eqnarray}\label{eq:NSEq}
\frac{\partial\left(\rho\boldsymbol{u}\right)}{\partial t} + \nabla\bcdot\left( \rho\boldsymbol{u}\otimes\boldsymbol{u}\right) & = & \nabla\bcdot\left[ \eta\left(\nabla\boldsymbol{u} + \nabla\boldsymbol{u}^T\right)\right] - \nabla p + \nabla p_{b,c} - C_3 \nabla\left(\frac{\delta f_{b,c}}{\delta C_3}\right)\nonumber\\
&& - \sum_{m,n = 1}^N \nabla\bcdot\left( c_{mn}\left(\nabla C_m\otimes\nabla C_n\right)\right). 
\end{eqnarray}

The equation of motion of the elastic membrane can be derived by considering the last four terms on the right-hand side of (\ref{eq:TDToralEntropy}). Setting their sum to zero and writing the equation for a discretized membrane for concreteness,  
\begin{equation}
\sum_{l = 1}^{N_l} \left[ \int_V  \frac{\delta f_{b,c}}{\delta\mathcal{I}}\frac{\partial\mathcal{I}}{\partial\boldsymbol{X}_l}
\frac{d\boldsymbol{X}_l}{dt} \ dV 
+ \frac{\partial\mathcal{E}_s}{\partial\boldsymbol{X}_l} \frac{d\boldsymbol{X}_l}{dt} 
+ \frac{\partial\mathcal{E}_b}{\partial\boldsymbol{X}_l} \frac{d\boldsymbol{X}_l }{dt} 
+ m \frac{d\boldsymbol{X}_l}{dt}\frac{d^2\boldsymbol{X}_l}{dt^2} \right] = 0.
\end{equation}
The variable $\boldsymbol{X}_l$ denotes the position of the $l^{th}$ Lagrangian marker (in Eulerian coordinates), with $l = 1,\dots,N_l$. Taking the expressions in the square bracket to be zero for each Lagrangian marker, we obtain
\begin{equation}\label{eq:MembraneForces}
m\frac{d^2\boldsymbol{X}_l}{dt^2} = \boldsymbol{F}_{mem,l} = \boldsymbol{F}_{s,l} + \boldsymbol{F}_{b,l} + \boldsymbol{F}_{c,l},
\end{equation}
where $\boldsymbol{F}_{mem}$ denotes the total force exerted on the elastic membrane, $\boldsymbol{F}_{s,l} = - \frac{\partial\mathcal{E}_s}{\partial \boldsymbol{X}_l}$ and $\boldsymbol{F}_{b,l} = - \frac{\partial\mathcal{E}_b}{\partial \boldsymbol{X}_l}$ are the strain and bending forces, and $\boldsymbol{F}_{c,l} = -\int_V \frac{\delta f_{b,c}}{\delta\mathcal{I}}\frac{\partial\mathcal{I}}{\partial\boldsymbol{X}_l}\ dV$ is the coupling force. The detailed forms of the forces are provided in \S\ref{subsec:MemDynamicsDiscr}.

\subsection{Free energy of the ternary fluid}\label{subsec:FluidFE}
As mentioned earlier, the free energy $F_f$ of the ternary fluid system considered here is the sum of two contributions: a Landau free-energy functional $\mathcal{E}_f$ allowing the coexistence of three fluid components, and a coupling-energy functional $\mathcal{E}_c$ taking into account the interaction between the membrane and its confined fluid component
\begin{equation}\label{eq:FluidFE}
F_f = \mathcal{E}_f + \mathcal{E}_c.
\end{equation}
In this work, we have used the following forms for $\mathcal{E}_f$ and $\mathcal{E}_c$
\begin{eqnarray}\label{eq:LandauFE}
\mathcal{E}_f & = & \int_V \left[ f_{b,f} + f_\nabla\right] dV \nonumber\\ & = & \sum_{m = 1}^N\int_V \left[ c_s^2\rho\ln\rho + \frac{\kappa_m}{2} C_m^2 \left( 1-C_m \right)^2 + \frac{\alpha^2\kappa_m}{2} \left(\nabla C_m\right)^2 \right] dV,
\end{eqnarray}
\begin{equation}\label{eq:CouplingE}
\hspace*{-110pt}\mathcal{E}_c = \int_V f_{b,c}\ dV = \int_V \frac{\kappa_c}{2}\left(C_3-\mathcal{I}\right)^2 dV.
\end{equation}
The first term on the right-hand side of (\ref{eq:LandauFE}) corresponds to the ideal gas term, as it is commonly used in multicomponent free energy lattice Boltzmann models \citep{KrugerEtAl2017}.

Each concentration fraction $C_m$ has two bulk minima at $C_m = \left\lbrace 0, 1\right\rbrace$. For the ternary fluid system of interest, only the following three minimizers are considered:
\begin{equation}\label{eq:Minimizers}
\begin{array}{ll}  
\displaystyle C_1 = 1,\ C_2 = 0,\ C_3 = 0;\\[8pt]
\displaystyle C_1 = 0,\ C_2 = 1,\ C_3 = 0;\\[8pt]
\displaystyle C_1 = 0,\ C_2 = 0,\ C_3 = 1, 
\end{array}
\end{equation}
corresponding to the three bulk fluids in the ternary system. As our multiphase fluid model assumes diffuse fluid-fluid interfaces, a formulation of the interfacial profile for the concentration of each component $m$ is required. Here it takes the form
\begin{equation}\label{eq:CmProfile}
C_m = \frac{1 + \tanh\left(d_{FI}/\left(2\alpha\right)\right)}{2},
\end{equation}
where $d_{FI}$ measures the distance between the bulk fluid at position $\boldsymbol{x}$ and the fluid-fluid interface. Equation (\ref{eq:CmProfile}) ensures that $C_m \rightarrow 1$ for $d_{FI} \rightarrow \infty$, and $C_m \rightarrow 0$ for $d_{FI} \rightarrow -\infty$. The parameter $\alpha$ is proportional to the interface width, which here is the same for all three fluid-fluid interfaces. The fluid-fluid surface tensions are expressed as 
\begin{equation}\label{eq:SurfaceTension}
\gamma_{mn} = \frac{\alpha}{6}\left(\kappa_m + \kappa_n\right),\ m,n = 1,2,3\ \mbox{and\ } m\neq n.
\end{equation}

The coupling energy is formulated in such a way that for $\kappa_c > 0$ there is a minimum when $C_3 = \mathcal{I}$. The interfacial profile $\mathcal{I}$ across the elastic membrane for the ternary fluid is then defined as 
\begin{equation}\label{eq:Iprofile}
\mathcal{I} = \frac{1+\tanh\left(d\left(\boldsymbol{x},\boldsymbol{X}\right) /\left( 2\alpha\right)\right)}{2},
\end{equation}
where $d\left(\boldsymbol{x},\boldsymbol{X}\right)$ denotes the distance between each bulk fluid component $m$ at position $\boldsymbol{x}$ and the membrane located at $\boldsymbol{X}$. This distance is assigned to be positive for the enclosed fluid component $3$, and negative for the surrounding fluid components $1$ and $2$. The width of the fluid-membrane interfaces is kept the same as the fluid-fluid interfaces one. It is obvious that with an increasing coupling coefficient $\kappa_c$, the interfacial profile for the concentration $C_3$ of the enclosed fluid component will be superimposed onto the profile of the fluid $3$-membrane interface. However, if $\kappa_c$ is too high, the coupling energy term will dominate over the rest of the free energy terms, which is undesirable. 

By comparing (\ref{eq:FEdensityGrad}) with the last term on the right-hand side of (\ref{eq:LandauFE}), it is obvious that
\begin{equation}\label{eq:Cmn}
c_{mn} = \alpha^2 \kappa_m\delta_{m,n},
\end{equation}
where $\delta_{m,n}$ is the Kronecker delta. Taking into account (\ref{eq:Cmn}), it can be shown that
\begin{equation}\label{eq:}
- \nabla p + \nabla p_{b,c} - C_3 \nabla\left(\frac{\delta f_{b,c}}{\delta C_3}\right) - \sum_{m,n = 1}^N \nabla\bcdot\left( c_{mn}\left(\nabla C_m\otimes\nabla C_n\right)\right) = -\sum_{m = 1}^N C_m \nabla\mu_m.
\end{equation}

\subsection{Coordinates Transformation}\label{subsec:CoordTransform}

For convenience, we shall express the equations of motion of the ternary fluid in terms of the mass density $\rho$, and the auxiliary fields $\phi$ and $\psi$, given by
\begin{equation}\label{eq:VariableTransform}
\rho = C_1 + C_2 + C_3,\ \phi = C_1 - C_2,\ \psi = C_3.
\end{equation}
Here we assume that all fluid components have the same density, and thus the mass density is set to be $\rho = 1$. Taking the aforementioned into account, the Navier-Stokes (\ref{eq:NSEq}) equations become
\begin{eqnarray}\label{eq:NSEqCT}
\frac{\partial\left(\rho\boldsymbol{u}\right)}{\partial t} + \nabla\bcdot\left( \rho\boldsymbol{u}\otimes\boldsymbol{u}\right) & = & \nabla\bcdot \eta\left(\nabla\boldsymbol{u} + \nabla\boldsymbol{u}^T\right) \nonumber\\
&& -\rho\nabla\mu_\rho - \phi\nabla\mu_\phi - \psi\nabla\mu_\psi - \psi\nabla \left(\frac{\delta f_{b,c}}{\delta\psi}\right), 
\end{eqnarray}
where $\mu_\rho$, $\mu_\phi$ and $\mu_\psi$ are the chemical potentials coming from the chosen free energy density of the ternary fluid. The detailed formulations of the chemical potentials $\mu_\rho$, $\mu_\phi$ and $\mu_\psi$ in terms of the variables $\rho$, $\phi$ and $\psi$ are given in the appendix \ref{appA}. The last term on the right-hand side of (\ref{eq:NSEqCT}) differentiates the present model from the ternary fluid model proposed by \citet{SemprebonEtAl2016}.

Equation (\ref{eq:MassLawComponent}) can be transformed into the Cahn-Hilliard equations governing the evolution of the order parameters $\phi$ and $\psi$ \citep{SemprebonEtAl2016}
\begin{equation}\label{eq:CHEqPhi}
\frac{\partial\phi}{\partial t} + \nabla\bcdot\left(\phi\boldsymbol{u}\right) = M_\phi\nabla^2 \mu_\phi,
\end{equation}
\begin{equation}\label{eq:CHEqPsi}
\frac{\partial\psi}{\partial t} + \nabla\bcdot\left(\psi\boldsymbol{u}\right) = M_\psi\nabla^2 \mu_\psi^\prime,
\end{equation}
where $M_\phi$ and $M_\psi$ are the mobility parameters, and $\mu_\psi^\prime = \mu_\psi + \frac{\delta f_{b,c}}{\delta\psi}$. Note that the corresponding diffusive term on the right-hand side of (\ref{eq:CHEqPsi}) in the ternary fluid model proposed by \citet{SemprebonEtAl2016} does not include the contribution owing to the coupling-energy functional, $\frac{\delta f_{b,c}}{\delta\psi}$. Here we assume that the variables $C_m \left( m = 1,2,3\right)$ have identical mobility parameters, resulting in $M_\phi = 3 M_\psi$. 

\section{Numerical methods}\label{sec:Numerics}
In this section, the numerical techniques employed for the solution of the equations of motion of the ternary fluid and its interacting elastic membrane are discussed. The governing equations of the ternary fluid are solved numerically by means of a lattice Boltzmann method, as detailed in \S\ref{subsec:LBM}. The discretized forms of the strain, bending and coupling energies are presented in \S\ref{subsec:MemDynamicsDiscr} along with the corresponding force formulations. The interaction between the ternary fluid and elastic membrane is solved by an immersed boundary method, described in \S\ref{subsec:IBM}, that is coupled to the lattice Boltzmann method following the algorithm presented in \citet{KrugerEtAl2011}. Finally, we briefly report the equivalent energies implementation in Surface Evolver, a software used for benchmarking purposes, in \S\ref{subsec:SE}.

To clarify notations, lower case letters are employed in the following for variables evaluated on the Eulerian lattices, while upper case ones refer to variables defined at the Lagrangian markers. 

\subsection{Lattice Boltzmann method}\label{subsec:LBM}
To solve the equations of motion of the ternary fluid (\ref{eq:ContEq}) and (\ref{eq:NSEqCT})--(\ref{eq:CHEqPsi}), we employ the lattice Boltzmann method with three sets of distribution functions $f_i\left(\boldsymbol{x}, t\right)$, $g_i\left(\boldsymbol{x},t\right)$, and $h_i\left(\boldsymbol{x}, t\right)$, corresponding to the fluid density $\rho$ and the order parameters $\phi$ and $\psi$. The evolutions of the distribution functions are governed by the lattice Boltzmann equation, where the standard Bhatnagar-Gross-Krook (BGK) single relaxation time model \citep{BhatnagarEtAl1954} is used for the collision operator, and the exact difference scheme \citep{Kupershtokh2004} is employed for the forcing term
\begin{eqnarray}\label{eq:LBE}
f_i \left(\boldsymbol{x}+\boldsymbol{c}_i\delta t, t+\delta t\right) & = & f_i \left(\boldsymbol{x}, t\right) - \frac{\delta t}{\tau}\left[ f_i\left( \boldsymbol{x}, t\right) - f_i^{eq}\left(\rho, \boldsymbol{u}\right)\right] \nonumber\\
&& + \delta t\left[f_i^{eq}\left(\rho,\boldsymbol{u}+\delta\boldsymbol{u}\right) - f_i^{eq}\left(\rho,\boldsymbol{u}\right)\right],\\[2pt]
g_i\left(\boldsymbol{x}+\boldsymbol{c}_i\delta t, t+\delta t\right) & = & g_i\left(\boldsymbol{x}, t\right) - \frac{\delta t}{\tau_\phi}\left[ g_i\left( \boldsymbol{x}, t\right) - g_i^{eq}\left(\phi,\boldsymbol{v}\right)\right],\\[2pt]
h_i\left(\boldsymbol{x}+\boldsymbol{c}_i\delta t, t+\delta t\right) & = & h_i\left(\boldsymbol{x}, t\right) - \frac{\delta t}{\tau_\psi}\left[ h_i\left(\boldsymbol{x}, t\right) - h_i^{eq}\left(\psi,\boldsymbol{v}\right) \right].
\end{eqnarray}

The variables $f_i\left(\boldsymbol{x}, t\right)$, $g_i\left(\boldsymbol{x}, t\right)$, and $h_i\left(\boldsymbol{x}, t\right)$ refer to the distribution functions $f_i$, $g_i$, and $h_i$ at position $\boldsymbol{x}$ and time $t$ with velocity $\boldsymbol{c}_i$ along the $i^{th}$ lattice direction. The relaxation times $\tau$, $\tau_\phi$ and $\tau_\psi$ are linked to the dynamic viscosity $\eta$, and the mobility parameters $M_\phi$ and $M_\psi$ by
\begin{eqnarray}\label{eq:RelaxationTimes}
\eta & = & \rho c_s^2\left(\tau-\frac{\delta t}{2}\right),\\[2pt]
M_\phi & = & \Gamma_\phi\left(\tau_\phi - \frac{\delta t}{2}\right),\\[2pt]
M_\psi & = & \Gamma_\psi\left(\tau_\psi - \frac{\delta t}{2}\right),
\end{eqnarray}
where $c_s$ is the speed of sound, and $\Gamma_\phi$, $\Gamma_\psi$ are tunable parameters. The speed of sound is given by $c_s = \frac{1}{\sqrt{3}} c$, where $c = \frac{\delta x}{\delta t}$ is the lattice speed, and $\delta x$, $\delta t$ are the lattice spacing and time step, respectively. The variables $f_i^{eq}$, $g_i^{eq}$, and $h_i^{eq}$ denote the equilibrium distribution functions.  

The latter are expressed as
\begin{eqnarray}\label{eq:EDF}
f_i^{eq}\left(\rho,\boldsymbol{u}\right) & = & w_i\rho\left[ 1 + \frac{\boldsymbol{c}_i\bcdot\boldsymbol{u}}{c_s^2} + \frac{\boldsymbol{u} \boldsymbol{u}\boldsymbol{:}\left(\boldsymbol{c}_i\boldsymbol{c}_i - c_s^2\mathsfbi{I}\right)}{2 c_s^4}\right],\\[5pt]
g_i^{eq}\left(\phi,\boldsymbol{v}\right) & = & \left\{
\begin{array}{ll}
w_i\left[\frac{\Gamma_\phi \mu_\phi}{c_s^2} + \frac{\phi\boldsymbol{c}_i\bcdot \boldsymbol{v}}{c_s^2} + \frac{\phi\boldsymbol{v}\boldsymbol{v}\boldsymbol{:} \left(\boldsymbol{c}_i\boldsymbol{c}_i-c_s^2\mathsfbi{I}\right)}{2 c_s^4}\right], & i\neq 0 \\[10pt]
\phi-\sum_{i, i\neq 0} g_i^{eq}, & i = 0,
\end{array}\right.\\[5pt]
h_i^{eq}\left(\psi,\boldsymbol{v}\right) & = & \left\{
\begin{array}{ll}
w_i\left[\frac{\Gamma_\psi \mu_\psi^\prime}{c_s^2} + \frac{\psi\boldsymbol{c}_i\bcdot \boldsymbol{v}}{c_s^2} + \frac{\psi\boldsymbol{v}\boldsymbol{v}\boldsymbol{:} \left(\boldsymbol{c}_i\boldsymbol{c}_i-c_s^2\mathsfbi{I}\right)}{2 c_s^4}\right], & i\neq 0 \\[10pt]
\psi-\sum_{i, i\neq 0} h_i^{eq}, & i = 0,
\end{array}\right.
\end{eqnarray}
where $w_i$ are weight coefficients depending on the chosen lattice arrangement for the velocity discretisation, and $\mathsfbi{I}$ is the identity tensor. As mentioned earlier, the chemical potential $\mu_\psi^\prime$ is given by $\mu_\psi^\prime = \mu_\psi + \frac{\delta f_{b,c}}{\delta\psi}$, where it is obvious from (\ref{eq:CouplingE}) that $\frac{\delta f_{b,c}}{\delta\psi} = \kappa_c\left(\psi-\mathcal{I}\right)$. The formulations of the chemical potentials $\mu_\phi$ and $\mu_\psi$ are provided in the appendix \ref{appA}.

The macroscopic physical variables are defined as moments of the distribution functions
\begin{eqnarray}\label{eq:MomentsDFs}
\rho\left(\boldsymbol{x}, t\right) & = & \sum_i f_i\left(\boldsymbol{x}, t\right),\\
\boldsymbol{u}\left(\boldsymbol{x}, t\right) & = & \left(\sum_i \boldsymbol{c}_i f_i\left(\boldsymbol{x}, t\right)\right)/\rho \left( \boldsymbol{x}, t\right),\\
\phi\left(\boldsymbol{x}, t\right) & = & \sum_i g_i\left(\boldsymbol{x}, t\right),\\
\psi\left(\boldsymbol{x}, t\right) & = & \sum_i h_i\left(\boldsymbol{x}, t\right).
\end{eqnarray}
The variable $\boldsymbol{u}$ represents the bare fluid velocity, and it is related to the actual fluid velocity $\boldsymbol{v}$ by
\begin{equation}\label{eq:ActualFluidVel}
\boldsymbol{v}\left(\boldsymbol{x}, t\right) = \boldsymbol{u}\left(\boldsymbol {x}, t\right) + \delta\boldsymbol{u}\left(\boldsymbol{x}, t\right)/2,
\end{equation}
where $\delta\boldsymbol{u}$ denotes the velocity correction given by
\begin{equation}\label{eq:VelCorrection}
\delta\boldsymbol{u}\left(\boldsymbol{x}, t\right) = \frac{\boldsymbol{f}\left( \boldsymbol{x}, t\right)}{\rho\left(\boldsymbol{x}, t\right)}\delta t.
\end{equation}

The forcing term $\boldsymbol{f}$ can be considered as the sum of three contributions: a force $\boldsymbol{f}_{FE}$ taking into account the gradient terms on the right-hand side of (\ref{eq:NSEqCT}), a force $\boldsymbol{f}_{IB}$ accounting for the interaction between the ternary fluid and elastic membrane, and a force $\boldsymbol{f}_{ext}$ allowing the existence of external forces
\begin{equation}\label{eq:ForcingTermLBE}
\boldsymbol{f} = \boldsymbol{f}_{FE} + \boldsymbol{f}_{IB} + \boldsymbol{f}_{ext}.
\end{equation}
The force $\boldsymbol{f}_{FE}$ can be written as
\begin{equation}\label{eq:ForceFE}
\boldsymbol{f}_{FE} = \nabla\bcdot\left( \rho c_s^2\mathsfbi{I} - \mathsfbi{p}\right) - \psi\nabla\left(\frac{\delta f_{b,c}}{\delta\psi}\right),
\end{equation}
where the pressure tensor $\mathsfbi{p}$ satisfies the condition 
\begin{equation}\label{eq:GradPT}
\nabla\bcdot\mathsfbi{p} = \rho\nabla\mu_\rho + \phi\nabla\mu_\phi + \psi\nabla\mu_\psi.
\end{equation}
The explicit form of $\mathsfbi{p}$ is provided in the appendix \ref{appA}. The form of the force $\boldsymbol{f}_{IB}$ is discussed in \S\ref{subsec:IBM}. In the present work, no external forces are considered, that is $\boldsymbol{f}_{ext} = \boldsymbol{0}$. 

In the current normalisation, the lattice spacing is set equal to the time step, $\delta x = \delta t = 1$, resulting in $c = 1$ and $c_s = 1/\sqrt{3}$. Both the relaxation times and the parameters $\Gamma_\phi$, $\Gamma_\psi$ are all considered to be equal to $1$, $\tau = \tau_\phi = \tau_\psi = 1$ and $\Gamma_\phi = \Gamma_\psi = 1$. Here, we employ the $D2Q9$ lattice arrangement, for which the lattice velocities are defined as
$$\boldsymbol{c}_i = c\left[
\begin{array}{rrrrrrrrr}
0 & 1 & -1 & 0 & 0 & 1 & -1 & -1 & 1\\\displaystyle
0 & 0 & 0 & 1 & -1 & 1 & -1 & 1 & -1\\
\end{array}
\right],$$
and the weight coefficients are given by
$$w_0 = 4/9,\ w_{1-4} = 1/9,\ w_{5-8} = 1/36.$$

\subsection{Membrane dynamics}\label{subsec:MemDynamicsDiscr}

The discretised strain and bending energies of the elastic membrane, given respectively by (\ref{eq:StretchingE}) and (\ref{eq:BendingE}) in its continuous-space forms, are formulated as
\begin{equation}\label{eq:DiscreteSE}
\mathcal{E}_s = \frac{\kappa_s}{2} \sum_{l = 1}^{N_l}\left(\frac{|\boldsymbol{X}_{l+1} - \boldsymbol{X}_l|}{\Delta s} - 1\right)^2 \Delta s,
\end{equation}
\begin{equation}\label{eq:DiscreteBE}
\mathcal{E}_b = \frac{\kappa_b}{4} \sum_{l = 1}^{N_l} \frac{\left(\Delta\theta \right)^2}{\Delta s^\prime}.
\end{equation}
The summation occurs over all the Lagrangian markers $l = 1,\dots,N_l$. The variable $\boldsymbol{X}_l$ denotes the position of the $l^{th}$ Lagrangian marker (in Eulerian coordinates), and $\Delta s$ represents the initial distance between two consecutive Lagrangian markers. The membrane is initially discretized into Lagrangian markers such that $\Delta s = \delta x= 1$. The tangential angle $\Delta\theta$ and arc length $\Delta s^\prime$ can be expressed as
\begin{equation}\label{eq:DiscreteTheta}
\Delta\theta\left( l\right) = 2\arccos \underbrace{\left[  \frac{\left(\boldsymbol{X}_{l+1}-\boldsymbol{X}_l \right)\bcdot\left(\boldsymbol{X}_l-\boldsymbol{X}_{l-1}\right)}{|\boldsymbol{X} _{l+1}-\boldsymbol{X}_l| |\boldsymbol{X}_l-\boldsymbol{X}_{l-1}|}\right] }_{y\left( l\right)},
\end{equation}
\begin{equation}\label{eq:DiscreteArcLength}
\Delta s^\prime\left( l\right) = |\boldsymbol{X}_l - \boldsymbol{X}_{l-1}| + |\boldsymbol{X} _{l+1} - \boldsymbol{X}_l|.
\end{equation}
Taking into account (\ref{eq:DiscreteSE})--(\ref{eq:DiscreteArcLength}), the corresponding discretised strain and bending forces can then be found as
\begin{eqnarray}\label{eq:DiscreteSF}
\boldsymbol{F}_{s} & = & \sum_{l = 1}^{N_l} \boldsymbol{F}_{s, l}  = -\sum_{l = 1}^{N_l} \frac{\partial\mathcal{E}_s}{\partial\boldsymbol{X}_l}\nonumber\\ 
& = & -\sum_{l = 1}^{N_l}\left( \sum_{m = 1}^{N_l}\kappa_s\left(\frac{|\boldsymbol{X}_{m+1} - \boldsymbol{X}_m|}{\Delta s} - 1\right) \frac{\left(\boldsymbol{X}_{m+1} - \boldsymbol{X}_m \right)}{|\boldsymbol{X}_{m+1} - \boldsymbol{X}_m|} \left(\delta_{m+1,l} - \delta_{m,l}\right)\right)
\end{eqnarray} 
and
\begin{eqnarray}\label{eq:DiscreteBF}
\boldsymbol{F}_b & = & \sum_{l = 1}^{N_l} \boldsymbol{F}_{b,l} = -\sum_{l = 1}^{ N_l} \frac{\partial\mathcal{E}_b}{\partial\boldsymbol{X}_l}\nonumber\\
& = & -\sum_{l = 1}^{N_l}\left(\sum_{m = 1}^{N_l} \frac{\kappa_b}{4}\left[ 2\frac{\Delta\theta\left( m\right)}{\Delta s^\prime\left( m\right)} \frac{\partial\Delta\theta\left( m\right)}{\partial\boldsymbol{X}_l} - \left(\frac{\Delta\theta\left( m\right)}{\Delta s^\prime\left( m\right)}\right)^2 \frac{\partial\Delta s^\prime\left( m\right)}{\partial\boldsymbol{X}_l} \right]\right),
\end{eqnarray}
where 
\begin{eqnarray}
\frac{\partial\Delta\theta\left( m \right)}{\partial\boldsymbol{X}_l} & = & - \frac{2}{\sqrt{1 - y^2\left( m\right)}}\left[\frac{\left(\boldsymbol{X}_m - \boldsymbol{X}_{m-1}\right)}{|\boldsymbol{X}_{m+1} - \boldsymbol{X}_m| |\boldsymbol{X}_m - \boldsymbol{X}_{m-1}|}\left(\delta_{m+1, l} - \delta_{m, l} \right)\right.\nonumber\\
&& \left. -\frac{\left(\boldsymbol{X}_{m+1} - \boldsymbol{X}_m\right)\bcdot \left(\boldsymbol{X}_m - \boldsymbol{X}_{m-1}\right)}{|\boldsymbol{X}_{m+1} - \boldsymbol{X}_m|^3 |\boldsymbol{X}_m - \boldsymbol{X}_{m-1}|} \left(\boldsymbol{X}_{m+1} - \boldsymbol{X}_m\right)\left(\delta_{m+1, l} - \delta_{m,l}\right)\right.\nonumber\\
&& \left. + \frac{\left(\boldsymbol{X}_{m+1} - \boldsymbol{X}_m\right)}{|\boldsymbol{X}_{m+1} - \boldsymbol{X}_m| |\boldsymbol{X}_m - \boldsymbol{X}_{m-1}|} \left(\delta_{m, l} - \delta_{m-1, l}\right)\right. \nonumber\\
&& \left. - \frac{\left(\boldsymbol{X}_{m+1} - \boldsymbol{X}_m\right)\bcdot \left(\boldsymbol{X}_m - \boldsymbol{X}_{m-1}\right)}{|\boldsymbol{X}_{m+1} - \boldsymbol{X}_m| |\boldsymbol{X}_m - \boldsymbol{X}_{m-1}|^3} \left( \boldsymbol{X}_m - \boldsymbol{X}_{m-1}\right) \left(\delta_{m, l} - \delta_{m-1, l}\right) \right], \nonumber
\end{eqnarray}
\begin{eqnarray}
\frac{\partial\Delta s^\prime\left( m\right)}{\partial\boldsymbol{X}_l} & = &  \frac{\left(\boldsymbol{X}_m - \boldsymbol{X}_{m-1}\right)}{|\boldsymbol{X}_m - \boldsymbol{X}_{m-1}|} \left(\delta_{m, l} - \delta_{m-1, l}\right) + \frac{\left(\boldsymbol{X}_{m+1} - \boldsymbol{X}_m\right)}{|\boldsymbol{X}_{m+1} - \boldsymbol{X}_m|} \left(\delta_{m+1, l} - \delta_{m, l}\right).\nonumber
\end{eqnarray}
The $\delta_{m, l}$ is the Kronecker delta. The variables $\boldsymbol{F}_{s, l}$ and $\boldsymbol{F}_{b, l}$ denote the discretised strain and bending forces exerted on the $l^{th}$ Lagrangian marker.

The discretised coupling energy of the elastic membrane can be written as
\begin{equation}\label{eq:DiscreteCE}
\mathcal{E}_c = \frac{\kappa_c}{2} \sum_{\boldsymbol{x}} \left(\psi - \mathcal{I}\right)^2 \left(\delta x\right)^d, 
\end{equation}
where $\sum_{\boldsymbol{x}}$ implies summation over all the Eulerian lattices $\boldsymbol{x}$, and $d$ is the domain dimensionality. The interfacial profile $\mathcal{I}$ is given by (\ref{eq:Iprofile}). The corresponding discretised coupling force can then be obtained as
\begin{eqnarray}\label{eq:DiscreteCF}
\boldsymbol{F}_c & = & \sum_{l = 1}^{N_l} \boldsymbol{F}_{c,l} = -\sum_{l = 1}^{N_l} \left(\sum_{\boldsymbol{x}}\frac{\delta f_{b,c}}{\delta\mathcal{I}} \frac{\partial\mathcal{I}}{\partial\boldsymbol{X}_l}\right) \nonumber\\ & = & -\sum_{l = 1}^{N_l} \left(\sum_{\boldsymbol{x}} \frac{\kappa_c}{4\alpha} \left(\psi-\mathcal{I}\right) \textsf{sech} ^2 \left(\frac{d\left(\boldsymbol{x}, \boldsymbol{X}_l\right)}{2\alpha} \right) \frac{\partial d\left(\boldsymbol{x}, \boldsymbol{X}_l\right)}{\partial\boldsymbol{X}_l} \right).
\end{eqnarray}
The variable $\boldsymbol{F}_{c, l}$ represents the discretised coupling force exerted on the $l^{th}$ Lagrangian marker. 

The discretised total force acting on the elastic membrane can thus be written as 
\begin{equation}\label{eq:DiscreteTF}
\boldsymbol{F}_{mem} = \sum_{l = 1}^{N_l} \boldsymbol{F}_{mem, l} = \sum_{l = 1}^{N_l} \left(\boldsymbol{F}_{s, l} + \boldsymbol{F}_{b, l} + \boldsymbol{F}_{c, l}\right).
\end{equation}
In the computational implementation, we exert the forces $\boldsymbol{F}_{s,l}$,    $\boldsymbol{F}_{b,l}$ and $\boldsymbol{F}_{c,l}$ on each Lagrangian marker $l$.

\subsection{Immersed boundary method}\label{subsec:IBM}

To reproduce the effect of $\boldsymbol{F}_{mem,l}$ on the Eulerian fluid flow, denoted here by $\boldsymbol{f}_{IB}$, a spreading operation is used 
\begin{equation}\label{eq:Spreading}
\boldsymbol{f}_{IB}\left(\boldsymbol{x}, t\right) = \mathcal{S}\left[ \boldsymbol{F}_{mem, l} \right]\left(\boldsymbol{x}\right) = \sum_{l = 1}^{N_l} \boldsymbol{F}_{mem, l}\delta_h\left(\boldsymbol{x} - \boldsymbol{X}_l\right) \Delta s.
\end{equation} 
The term $\delta_h$ denotes the discretised Dirac delta function, and the following formulation proposed by \citet{Peskin2002} is chosen here to perform the convolution in the spreading and interpolation operations
\begin{eqnarray}\label{eq:DiscreteDirac}
\delta_h\left( r\right) & = & \left\{
\begin{array}{ll}
\frac{1}{8}\left( 3 - 2 |r| + \sqrt{1 + 4 |r| - 4 r^2}\right), & |r|\leq 1 \\[10pt]
\frac{1}{8}\left( 5 - 2 |r| - \sqrt{-7 + 12 |r| - 4 r^2}\right), & 1\leq |r|\leq 2\\[10pt]
0, & 2\leq |r|.
\end{array}\right.
\end{eqnarray}
If the variable r is a vector, $\boldsymbol{r} = \left( r_x, r_y, r_z\right)$, then the multidimensional $\delta_h$ is given by $\delta_h\left(\boldsymbol{r} \right) = \frac{1}{\delta x^3}\delta_h\left(\frac{r_x}{\delta x}\right) \delta_h\left(\frac{r_y}{\delta x}\right) \delta_h\left(\frac{r_z}{\delta x}\right)$. Once the force $\boldsymbol{f}_{IB}$ is computed, the density $\rho$, order parameters $\phi$ and $\psi$, and velocity $\boldsymbol{v}$ fields can be obtained at the next time step $t + \delta t$ by solving (\ref{eq:LBE}). To calculate the forces $\boldsymbol{F}_{mem,l}$ at $t + \delta t$, the position of the Lagrangian markers $\boldsymbol{X}_l,\ l = 1,\dots, N_l$ at $t+\delta t$ needs to be known. For this reason, the known $\boldsymbol{v}\left( \boldsymbol{x}, t+\delta t\right)$ is interpolated at the Lagrangian markers as
\begin{equation}\label{eq:Interpolation}
\boldsymbol{U}\left(\boldsymbol{X}_l, t+\delta t\right) = \mathcal{I}\left[ \boldsymbol{v}\right]\left(\boldsymbol{X}_l\right) = \sum_{\boldsymbol{x}} \boldsymbol{v} \delta_h\left(\boldsymbol{x}-\boldsymbol{X}_l\right) \left( \delta x\right)^d. 
\end{equation} 
The updated position of the Lagrangian markers can then be found by Euler's rule
\begin{equation}\label{eq:UpdatedXl}
\boldsymbol{X}_l\left( t+\delta t\right) = \boldsymbol{X}_l\left( t\right) + \boldsymbol{U}\left(\boldsymbol{X}_l, t+\delta t\right) \delta t.
\end{equation}

\subsection{Surface Evolver}\label{subsec:SE}

In the absence of closed-form solutions for non-trivial elastocapillary problems, we have benchmarked the proposed dynamic simulation method against a finite element approach. In particular, the shapes of elastic capsules placed at a fluid-fluid interface in mechanical equilibrium have been modelled. For the finite element approach we employ Surface Evolver, an open-source software developed by \citet{Brakke1992}, which has been extensively used to model the equilibrium shapes of liquid interfaces and capillary forces \citep{GrzybowskiEtAl2001,Collicott2004,SemprebonEtAl2016b}, detect energy barriers of morphological transitions \citep{BiEtAl2014, SemprebonEtAl2018}, and model deformations of elastic membranes  \citep{ManyuhinaEtAl2010}.

In Surface Evolver the interfaces are discretized by triangulated meshes constituted of vertices, edges and facets. Surface Evolver provides a large library of pre-implemented energy functionals, including surface tension, elastic strain and bending energies, from which the forces acting on each vertex are automatically computed within the internal routines. Local and global constraints can also be implemented in Surface Evolver, such as to maintain volume/area conservation and to keep certain vertices fixed. Configurations in mechanical equilibrium correspond to minima of the total energy, obtained through a conjugate gradient descent method.

For our benchmarks, we employ a 2D model in Surface Evolver, where the elastic capsules are initialized with exactly the same geometry as in the proposed coupled lattice Boltzmann-immersed boundary method (LB-IBM). The same free energy of the elastic membrane is implemented using the provided scripting language to formulate the strain and bending energies as in (\ref{eq:DiscreteSE}) and (\ref{eq:DiscreteBE}). In particular, the functions \texttt{edge\_length} and \texttt{sqcurve\_string\_marked} are employed for the calculation of the strain and bending energies. The main difference with the LB-IBM is in the definition of the surface tensions; here, the corresponding energy is simply accounted for by adding a term proportional to the total length of each fluid or membrane interface, multiplied by a constant parameter matching the surface tension arising from the diffuse interface in the lattice Boltzmann method. No coupling energy has been considered, as both the elastic energies and surface tension are provided by the discrete representation of the capsule and the fluid-fluid interface. The constraint of conservation of the total capsule area has also been considered. Finally, the position of the vertices at the end of the computational domain has been kept fixed throughout the minimization procedure, as an equivalent to implementing periodic boundary conditions in the LB-IBM.

\section{Results}\label{sec:Results}
The proposed fluid-structure solver is validated in \S\ref{subsec:LiquidLens}. The steady configuration of an elastic capsule positioned at a fluid-fluid interface is chosen for this purpose. We perform a thorough comparison with the reference results of Surface Evolver for different cases of surface tension ratios, and various combinations of Young's and bending moduli. We also establish the Galilean invariance of the governing equations of the ternary fluid and elastic structure. Finally, a more complex configuration is considered in \S\ref{subsec:CapillaryBridges} to demonstrate the capabilities of the proposed model.

\subsection{Elastic capsule at fluid-fluid interface}\label{subsec:LiquidLens}
\begin{figure}
\centerline{
\includegraphics[scale=0.35]{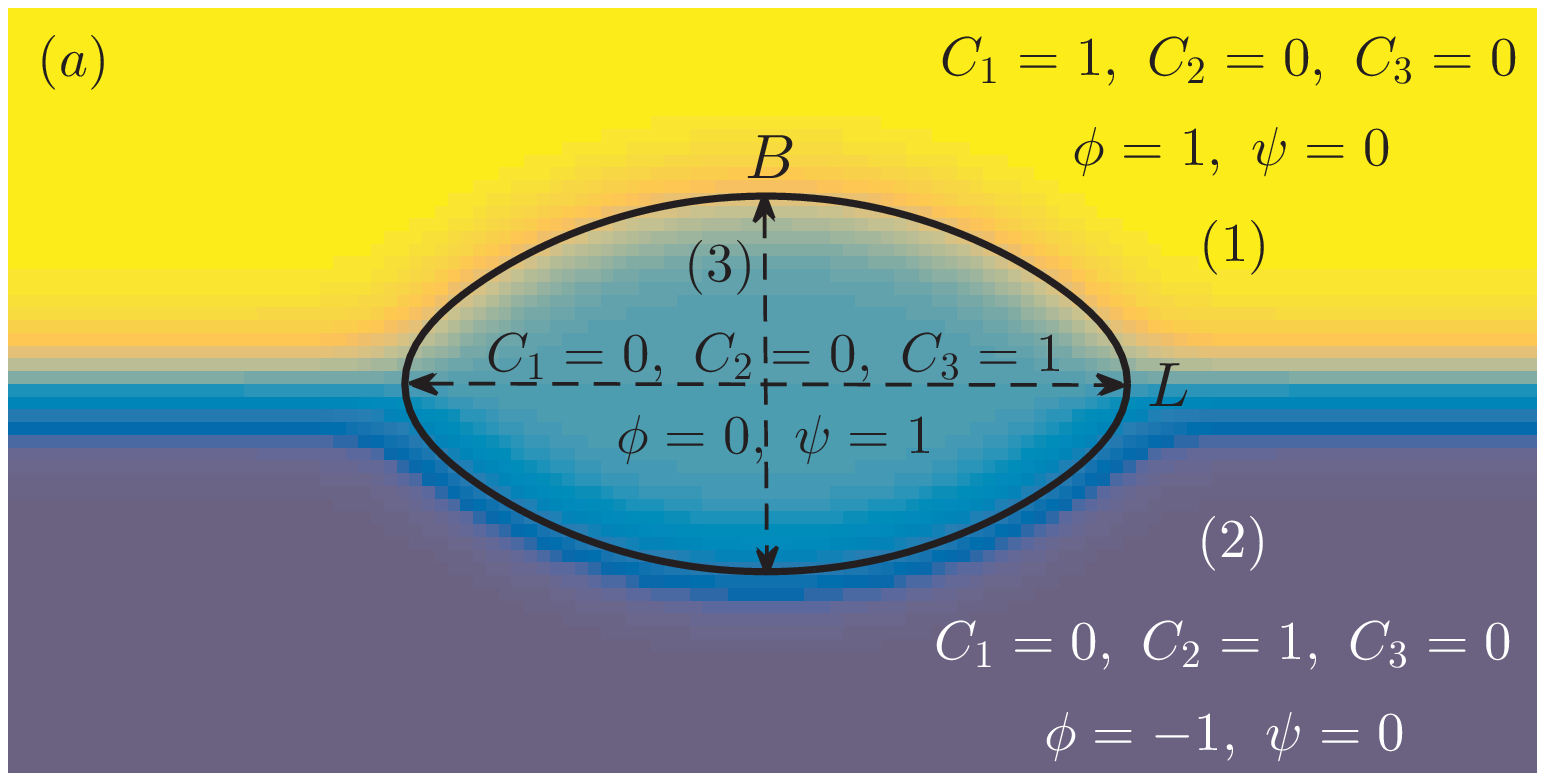}
\includegraphics[scale=0.355]{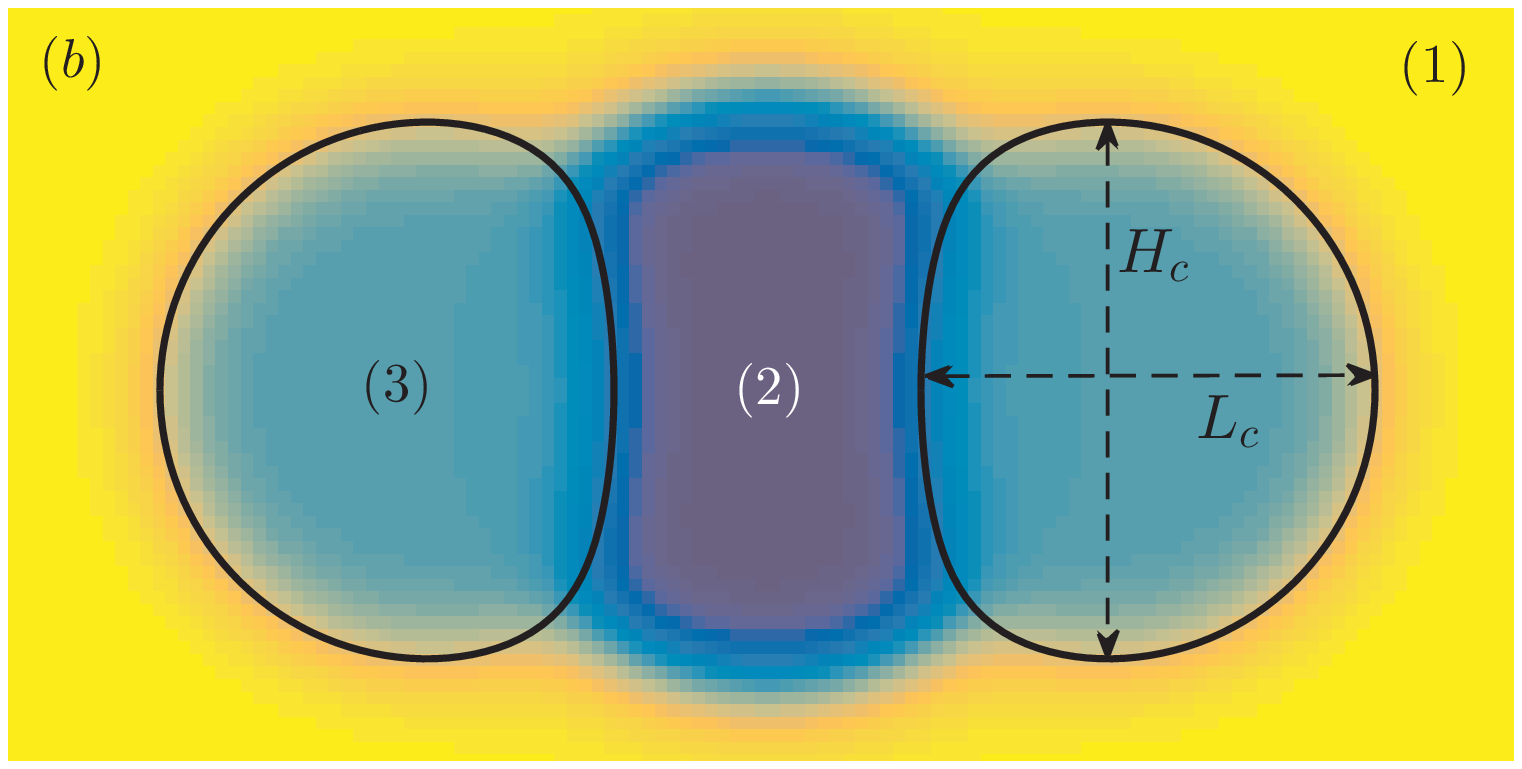}
}
\caption{Schematic diagrams for: $\left( a\right)$ an elastic capsule at a fluid-fluid interface, and $\left( b\right)$ a capillary bridge formed between two elastic capsules. The figures are simulation snapshots focused in the area around the elastic membranes, where diffuse interfaces of width $\alpha = 2$ are considered. The capsules (---) are filled with the fluid component $3$, while being surrounded by the fluid components $1$ and $2$.}
\label{fig:Figure1}
\end{figure}

To benchmark the proposed model, the configuration of an elastic capsule placed at a fluid-fluid interface, as shown in figure \ref{fig:Figure1}$\left( a\right)$, is studied. An initially circular capsule of radius $R = 20$ is located at the center of a computational domain of dimensions $12 R \times 6 R$. The capsule relaxes to a mechanical equilibrium shape, which depends on the balance of the elastic strain and bending forces and the surface tensions $\gamma_{12}$, $\gamma_{13}$ and $\gamma_{23}$. We can define a dimensionless deformation parameter, the Taylor deformation $D$, given by $D = \left( L-B\right) / \left( L+B\right)$, where $L$ and $B$ are, respectively, the major and minor axes of the final elliptical capsule shape. Periodic boundary conditions are considered at all the domain boundaries. Simulations are performed for several combinations of Young's and bending moduli, $\kappa_s = \left\lbrace 10^{-3}, 10^{-2}, 10^{-1} \right\rbrace$ and $\kappa_b = \left\lbrace 10^{-4}, 10^{-2}\right\rbrace$, and various surface tensions. The parameter $\alpha$ in (\ref{eq:SurfaceTension}) is set to be $\alpha = 2$. The coupling coefficient is taken as $\kappa_c = 10^{-2}$, which was found to be the threshold before the domination of the coupling energy over the rest of the free energy terms occurs. This value of the coupling coefficient also ensured that the interfacial profile for the concentration $C_3$ of the enclosed fluid component is superimposed onto the profile of the fluid 3-membrane interface.  

\begin{figure}
\centerline{
\includegraphics[scale=0.35]{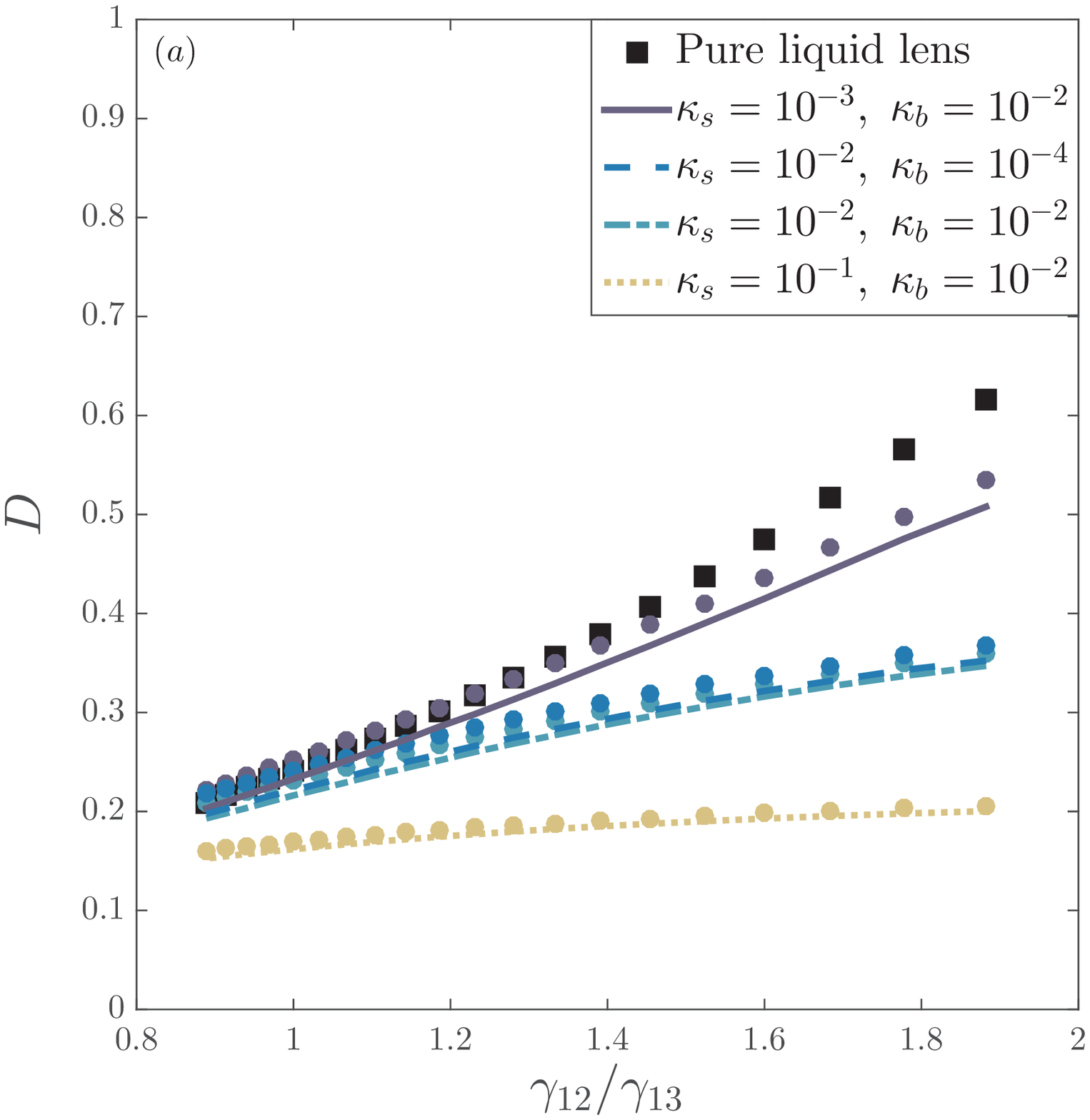}
\includegraphics[scale=0.35]{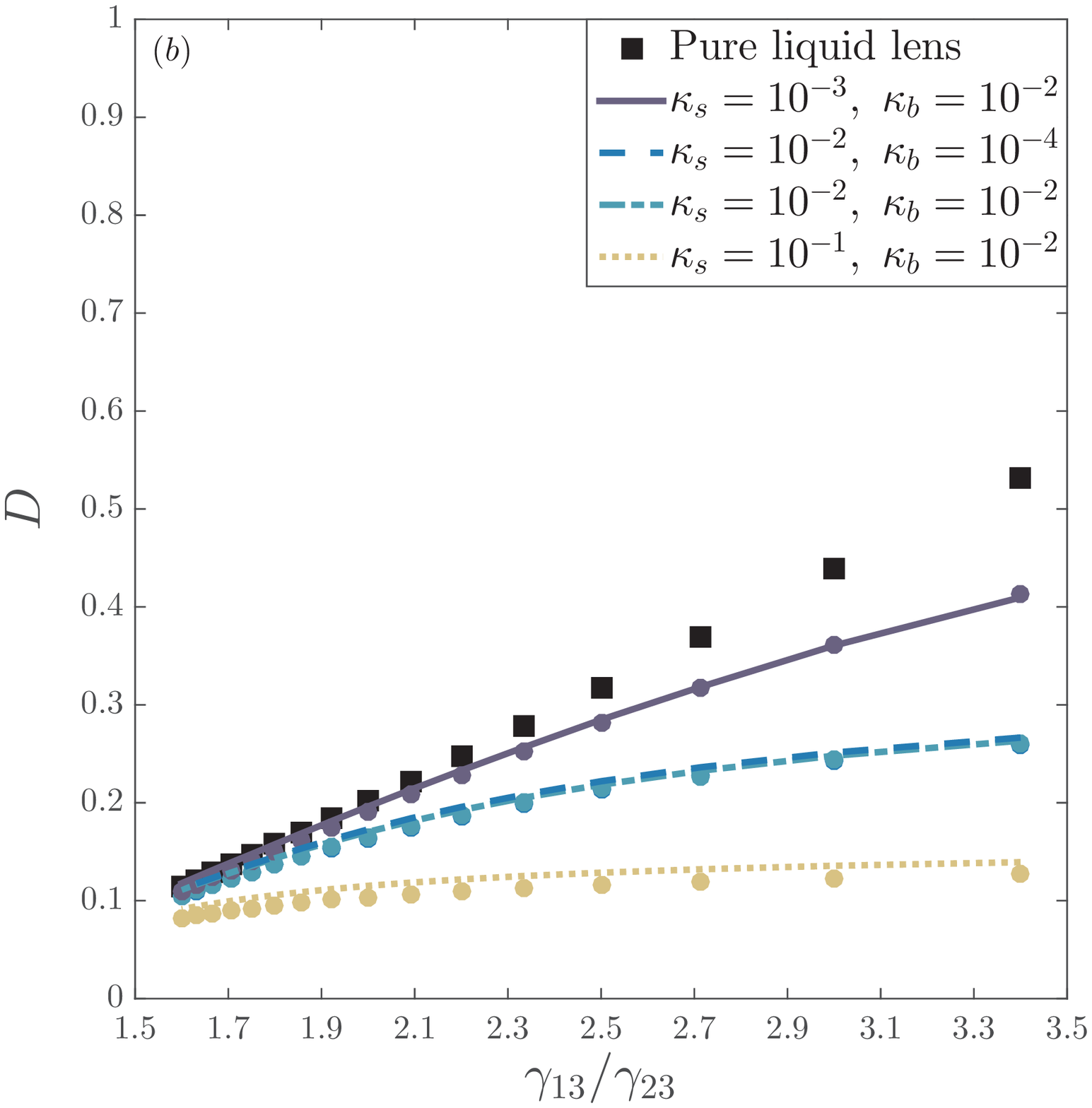}
}
\caption{Comparison of the Taylor deformation $D$ between the results of the fluid-structure solver and Surface Evolver (\textbullet) for: $\left( a\right)$ the symmetric case, and $\left( b\right)$ the asymmetric one. The results of the fluid solver for the pure liquid lens configuration are also presented ($\blacksquare$).}
\label{fig:Figure2}
\end{figure}

We first consider the case where $\gamma_{13} = \gamma_{23}$, resulting in a capsule shape that is symmetrical along the domain centreline. We will refer thereafter to this case as the symmetric case. For the surface tensions $\gamma_{13}$ and $\gamma_{23}$ to be equal, the values of the parameters $\kappa_1$ and $\kappa_2$ in (\ref{eq:SurfaceTension}) have to be the same, taken here as $\kappa_1 = \kappa_2 = 8\cdot 10^{-3}$. The parameter $\kappa_3$ ranges from $5\cdot 10^{-4}$ to $10^{-2}$ with a step size of $5\cdot 10^{-4}$. The corresponding surface tension ratio $\gamma_{12} / \gamma_{13} = \gamma_{12} / \gamma_{23}$ varies from $0.88\bar{8}$ to $1.882$. Figure \ref{fig:Figure2}$\left( a\right)$ shows the Taylor deformations obtained by our simulations, whose results are represented by lines, compared to the ones measured by Surface Evolver, denoted by dot symbols. As expected, the variations in the Taylor deformations found for a particular combination of $\kappa_s$ and $\kappa_b$ become more apparent with a decrease in the Young's modulus. By lowering the latter, the results for the elastic capsule tend to the ones obtained for the pure liquid lens configuration, depicted by filled square symbols. It is also obvious that the effect of the bending coefficient on the capsule's shape is negligible for a given Young's modulus. 

Our results generally agree well with those of Surface Evolver, with a typical relative error in $D$ of $5.6\%$. The discrepancies in the Taylor deformation between the two numerical methods increase for highly deformable capsules, that is $\kappa_s \leq 10^{-3}$, subjected to high surface tension ratios. We find that this is due to spurious forces at the three-phase contact points. An increase in the interface width $\alpha$ and the computational domain size could help in the damping of these spurious effects, with an increase, however, in the corresponding computational cost. In any case, the relative errors in $D$ between the results of our simulations and Surface Evolver are less than $10\%$, a reasonable limit given the significant differences in the two solution techniques.

\begin{figure}
\centerline{
\includegraphics[scale=0.35]{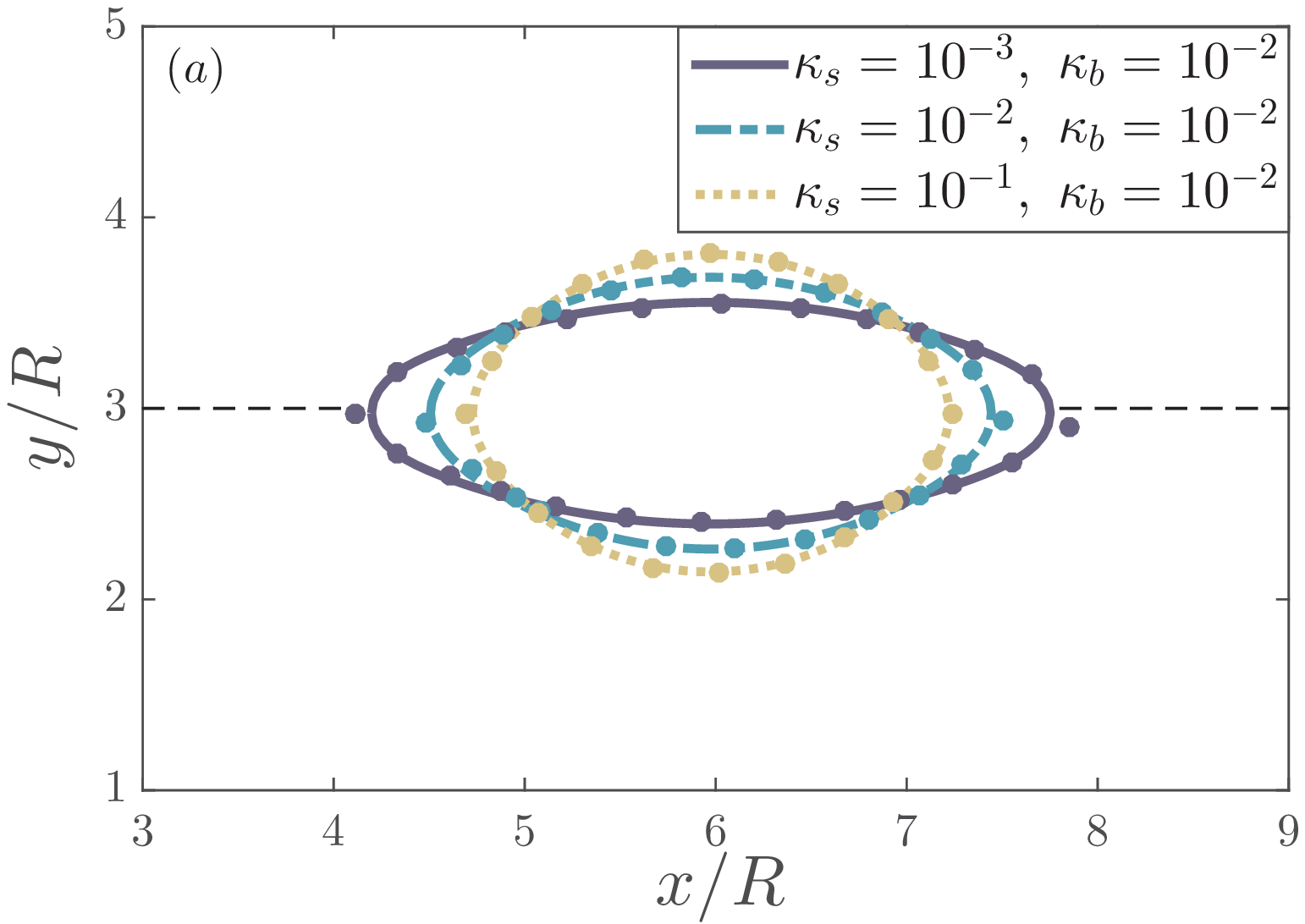}
\includegraphics[scale=0.35]{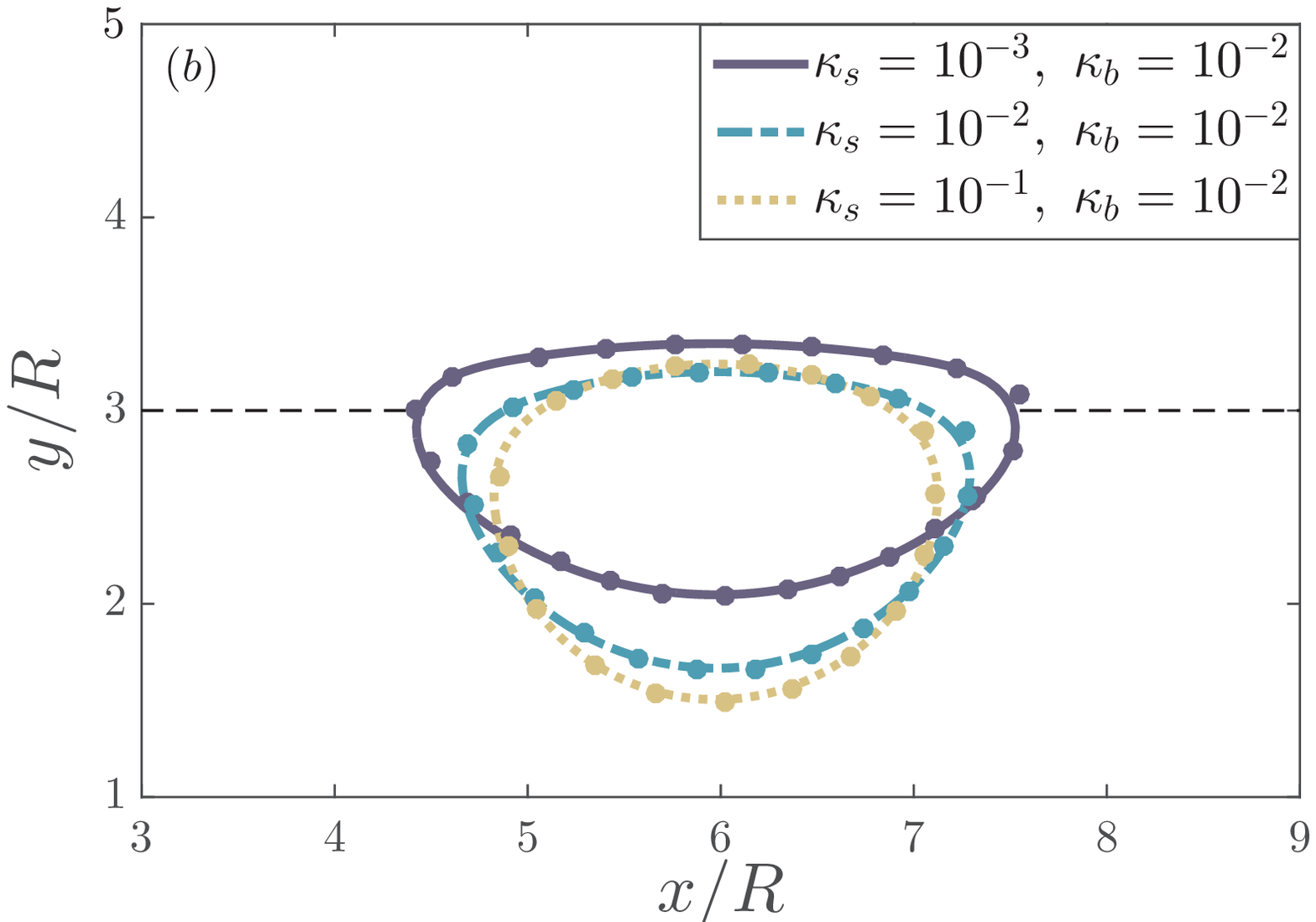}
}
\caption{Mechanical equilibrium shapes of the elastic capsule for the: $\left( a\right)$ symmetric, and $\left( b\right)$ asymmetric cases. The results of our simulations are compared to those of Surface Evolver (\textbullet), which are sub-sampled by a factor of $4$ for viewing clarity. The dash lines (- - -) represent the middle of the diffuse interface separating fluids $1$ and $2$.}
\label{fig:Figure3}
\end{figure}

To illustrate the effect of these discrepancies in the Taylor deformation on the mechanical equilibrium shape of the elastic capsule, the latter is plotted for different combinations of $\kappa_s$ and $\kappa_b$ at the highest surface tension ratio studied here, that is $\gamma_{12} / \gamma_{13} = 1.882$, as shown in figure \ref{fig:Figure3}$\left( a\right)$. For stiff and moderately deformable capsules, corresponding to $\kappa_s = 10^{-1}$ and $\kappa_s = 10^{-2}$, an excellent agreement is observed between our simulation results and those of Surface Evolver. For highly deformable capsules, a slightly elongated shape is obtained by the reference software compared to the one found by our fluid-structure solver. The relative error in $L$ is measured to be $5.75\%$. The capsule width agrees, however, perfectly.

We subsequently investigate the case where $\gamma_{13} \neq \gamma_{23}$, resulting in the capsule to be more immersed in one of the surrounding fluid phases. This case will be mentioned in the following as the asymmetric case. We assign the parameters $\kappa_1$ and $\kappa_2$ to be $\kappa_1 = 8\cdot 10^{-3}$ and $\kappa_2 = 2\cdot 10^{-3}$. The parameter $\kappa_3$ varies from $5\cdot 10^{-4}$ to $8\cdot 10^{-3}$ in steps of $5\cdot 10^{-4}$, with the corresponding surface tension ratio $\gamma_{13} / \gamma_{23}$ ranging from $1.60$ to $3.40$. The dependence of the measured Taylor deformations on the aforementioned surface tension ratio is presented for the LB-IBM and Surface Evolver results in figure \ref{fig:Figure2}$\left( b\right)$. Similar observations to the symmetric case can be made on the trends of $D$ for different Young's and bending moduli. The plots of the mechanical equilibrium capsule shapes are depicted for different combinations of $\kappa_s$ and $\kappa_b$ at the highest surface tension ratio achieved here, namely $\gamma_{13} / \gamma_{23} = 3.40$, in figure \ref{fig:Figure3}$\left( b\right)$. We can demonstrate again an excellent agreement between the results of our algorithm and Surface Evolver.

\begin{figure}
\centerline{
\includegraphics[scale=0.35]{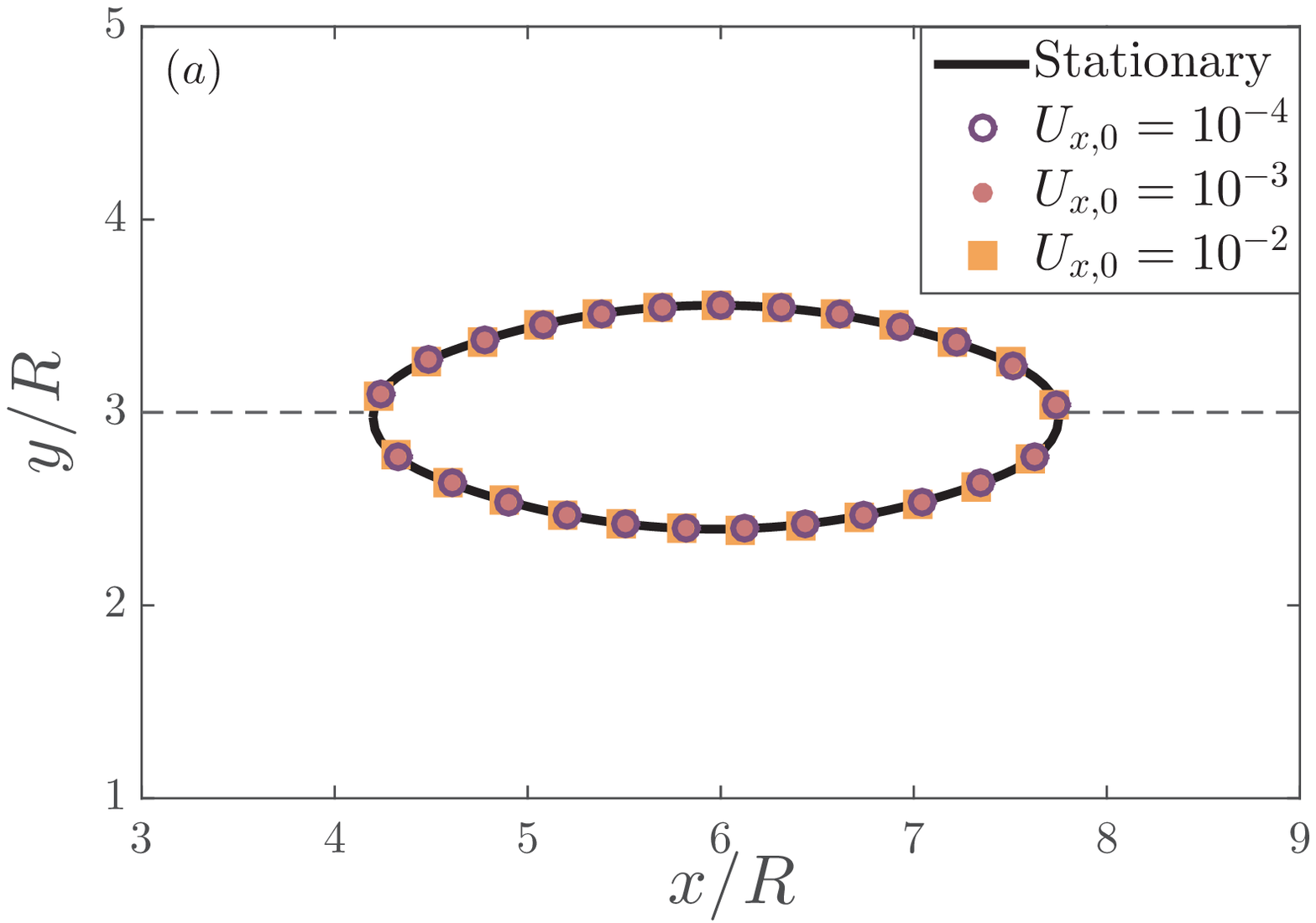}
\includegraphics[scale=0.35]{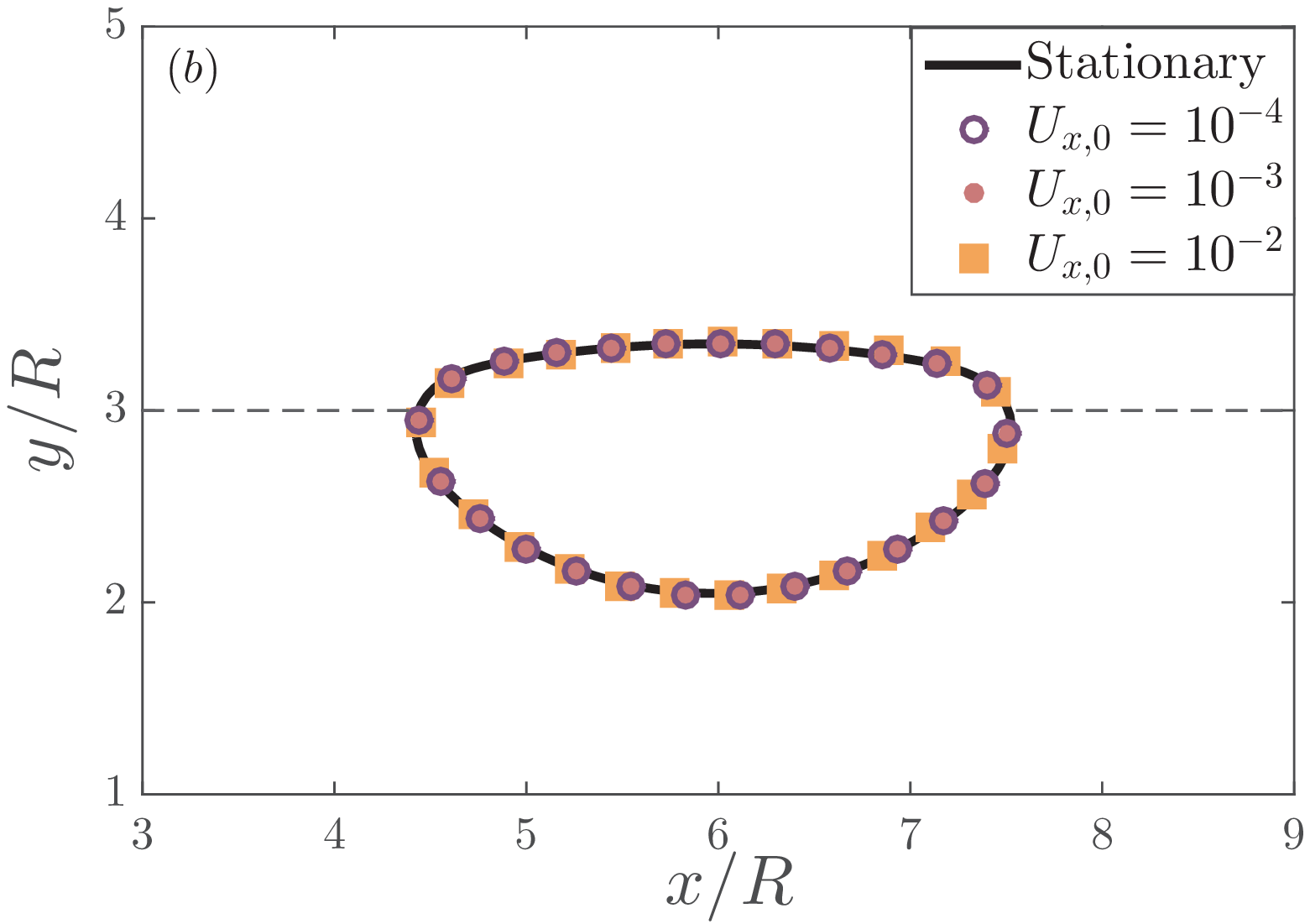}}
\caption{Comparison of the capsule shapes between the stationary and inertial reference frames for: $\left( a\right)$ the symmetric case, and $\left( b\right)$ the asymmetric one. For viewing clarity, the Lagrangian markers depicted here for the results in the inertial reference frame are sub-sampled by a factor of $5$. The middle of the diffuse interface separating fluids $1$ and $2$ is denoted by dash lines (- - -).}
\label{fig:Figure4}
\end{figure}

We finally prove the Galilean invariance of the proposed model. To do so, simulations are performed in an inertial frame of reference, and their results are compared to the ones obtained by the previous stationary simulations. At $t = 0$, all the fluid components are given a constant horizontal velocity $U_{x, 0}$. Due to the fluid-structure interaction, the initially circular capsule travels in the same direction with an equal velocity. Three different velocity values are tested here, $U_{x, 0} = \left\lbrace 10^{-4}, 10^{-3}, 10^{-2} \right\rbrace$. The Galilean invariance is checked in both the symmetric and asymmetric cases for $\kappa_s = 10^{-3}$ and $\kappa_b = 10^{-2}$ at the highest surface tension ratios examined respectively here. The shapes of the elastic capsule, after the latter reaches mechanical equilibrium in the inertial reference frame, found for the different velocities are compared to each other and to the ones of the corresponding stationary simulations. These comparisons can be seen in figure \ref{fig:Figure4}. In both cases, the results for the capsule shapes in the moving and stationary frames are perfectly superimposed, indicating that the proposed model is Galilean invariant.

\subsection{Capillary bridge between two elastic capsules}\label{subsec:CapillaryBridges}
To show the capabilities of our model, the configuration of a capillary bridge formed between two deformable capsules, as shown in figure \ref{fig:Figure1}$\left( b\right)$, is now investigated. Two initially circular capsules of radius $R = 20$ are placed at $\left( x_{c_1}, y_{c_1}\right) = \left( 9 R / 2, 3 R\right)$ and $\left( x_{c_2}, y_{c_2}\right) = \left( 15 R / 2, 3 R\right)$ of a computational domain of dimensions $12 R\times 6 R$. The capillary bridge, composed of the fluid component $2$, is initialized as a rectangular area of dimensions $2 S \times H_b = R\times 31 R / 10$ located at the center of the computational domain. Both the capillary bridge and elastic capsules are surrounded by the fluid component $1$. Due to the presence of the surface tensions $\gamma_{12}$, $\gamma_{13}$ and $\gamma_{23}$ as well as the elastic strain and bending forces, the capsules  relax to a deformed mechanical equilibrium shape depending on the balance of the aforementioned forces. This mechanical equilibrium shape depends also on the initial distance between the two capsules, $2 S$, and the volume of the capillary bridge. To confine the parameter space of the current study, all the simulations are performed only for a dimensionless initial distance $S^\prime = 2 S / R = 1$, and a relative area $A_{rel} = A_{b} / A_{c} \approx 1.02$, where $A_b$ and $A_c$ are the areas of the capillary bridge and elastic capsules. Three different surface tension ratios are tested here, $\gamma_{12} / \gamma_{13} \approx \left\lbrace 0.57, 0.67, 0.80 \right\rbrace$. The respective parameters in (\ref{eq:SurfaceTension}) are set to be $\alpha = 2$, $\kappa_1 = \left\lbrace 1\cdot 10^{-3}, 3\cdot 10^{-3}, 9\cdot 10^{-3}\right\rbrace$, $\kappa_2 = 3\cdot 10^{-3}$ and $\kappa_3 = 6\cdot 10^{-3}$. The simulations are performed for various Young's moduli, $\kappa_s = \left\lbrace 10^{-4}, 10^{-3}, 10^{-2}, 10^{-1}, 10^0 \right\rbrace$, and constant bending and coupling coefficients, $\kappa_b = 10^{-2}$ and $\kappa_c = 10^{-2}$. Periodic boundary conditions are applied at all the domain boundaries. At the converged state, the dimensionless aspect ratio of the elastic capsules can be defined as $L_c / H_c$.

\begin{figure}
\centerline{\includegraphics[scale=0.35]{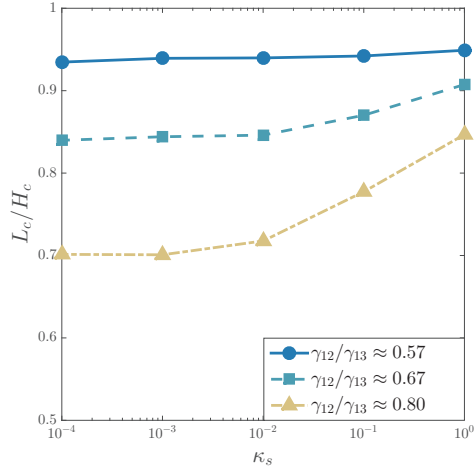}}
\caption{Aspect ratio $L_c / H_c$ of the elastic capsules as a function of  Young's modulus $\kappa_s$ at different surface tension ratios $\gamma_{12} / \gamma_{13}$.}
\label{fig:Figure5}
\end{figure}

\begin{figure}
\centerline{
\includegraphics[scale=0.35]{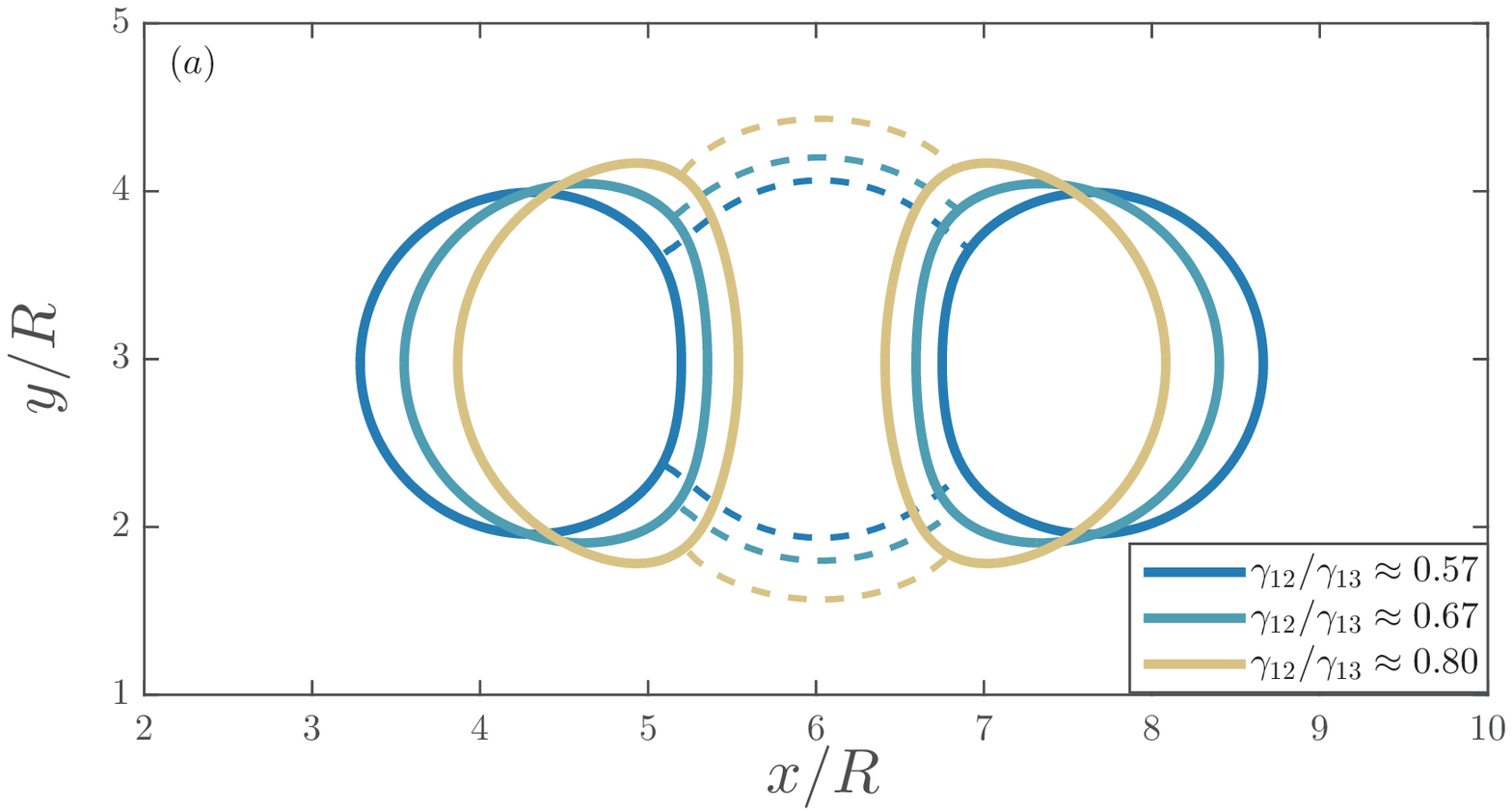}
\includegraphics[scale=0.35]{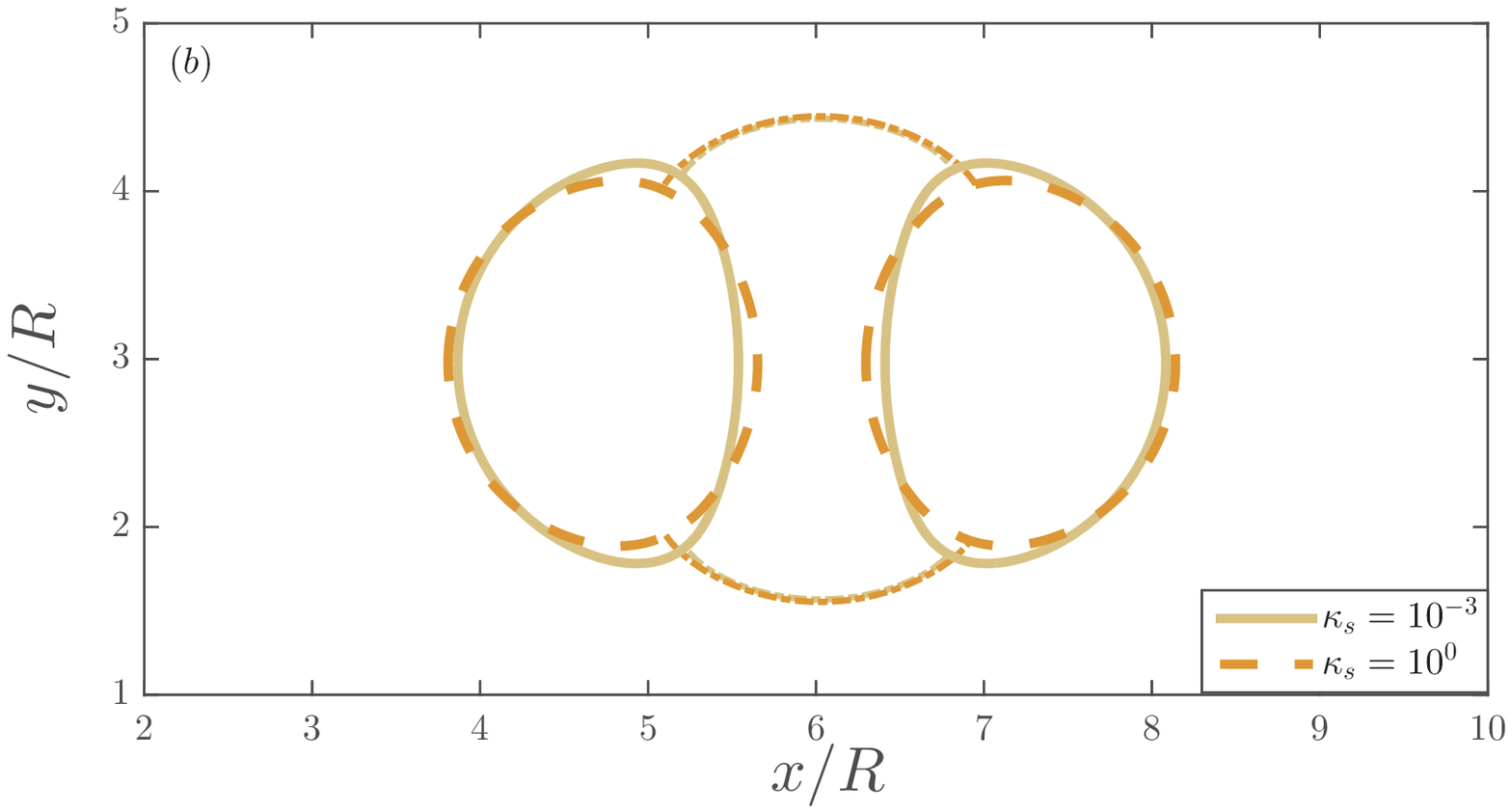}
}
\caption{Comparison of the mechanical equilibrium shapes of the elastic capsules for: $\left( a\right)$ different surface tension ratios $\gamma_{12} / \gamma_{13}$ at $\kappa_s = 10^{-3}$, and $\left( b\right)$ different Young's moduli $\kappa_s$ at $\gamma_{12} / \gamma_{13} \approx 0.80$. The dash (- - -) and dash-dot (-.) lines in $\left( a\right)$ and $\left( b\right)$, respectively, depict the capillary bridge boundaries.}
\label{fig:Figure6}
\end{figure}

The effect of Young's modulus on the aspect ratio of the capsules is presented in figure \ref{fig:Figure5} for the different surface tension ratios. For low $\gamma_{12} / \gamma_{13}$, the capsules seem to take similar mechanical equilibrium shapes independently of their degree of elasticity. For moderate and high $\gamma_{12} / \gamma_{13}$, the aspect ratio changes significantly between stiff and moderately deformable $\left(\kappa_s = 10^{-2}\right)$ capsules, while it reaches a plateau for highly deformable $\left(\kappa_s \leq 10^{-3} \right)$ capsules. It can also be noted that the surface tension ratio affects considerably the capsules' aspect ratio for a given Young's modulus. This can be clearly seen in figure \ref{fig:Figure6}$\left( a\right)$, where the mechanical equilibrium shapes of the highly deformable $\left(\kappa_s = 10^{-3}\right)$ capsules are plotted for the different $\gamma_{12} / \gamma_{13}$. At $\gamma_{12} / \gamma_{13} \approx 0.57$, the capsules retain an almost circular shape, having only the part of their surface coming in contact with the capillary bridge slightly compressed. As the surface tension ratio increases, the capsules move towards each other, taking a semi-circular shape and causing the formation of a narrower and higher capillary bridge. Despite the fact that Young's modulus has a notable effect on the aspect ratio of the elastic capsules at high $\gamma_{12} / \gamma_{13}$, the corresponding variations in their shapes are small for different $\kappa_s$, as shown in figure \ref{fig:Figure6}$\left( b\right)$. The capsule becomes slightly shorter and wider with an increase in $\kappa_s$. 

\begin{figure}
\centerline{
\includegraphics[scale=0.35]{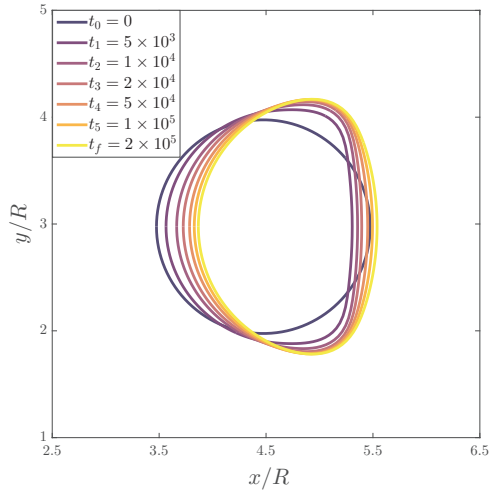}
}
\caption{Transient shapes of the highly $\left(\kappa_s = 10^{-3}\right)$ deformable capsule at $\gamma_{12}/\gamma_{13} \approx 0.80$.}
\label{fig:Figure7}
\end{figure}

Finally, the transient shapes of the highly deformable capsules at a surface tension ratio $\gamma_{12}/\gamma_{13} \approx 0.80$ are shown in figure \ref{fig:Figure7}. For clarity, we present only the results of the left-hand side capsule; the transient shapes of the right-hand side capsule are symmetric to the ones presented across the vertical centreline. As mentioned previously, the elastic capsules have initially, at $t_0 = 0$, a circular shape. It is worth noting that the capsules take quickly, already at $t = t_1$, a semi-circular shape similar to the mechanical equilibrium one. As the time passes, the capsules become narrower and more elongated along the $x-$ and $y-$axis, respectively. Minimal changes in the capsules shapes can be observed between $t_5$, which corresponds to half of the total simulation time, and $t_f$, where the capsules are considered to have reached the mechanical equilibrium shape.

\section{Conclusions}\label{sec:Conclusions}
In this work, we have presented a novel simulation technique for the coupling of a multicomponent fluid flow with deformable, infinitely thin structures. For the sake of simplicity, we have considered the case where elastic membranes enclosing a fluid component $3$ are immersed in a two-component/phase flow consisted of fluids $1$ and $2$. For this case, we have introduced a modified formulation of the free energy of the ternary fluid, taking into consideration its interaction with the elastic membranes, to the original one proposed by \citet{SemprebonEtAl2016}. Having as a starting point an equation for the conservation of the entropy, we have made use of the splitting structure of the latter and thermodynamic laws in order to derive the equations governing the evolution of the ternary fluid. Following a similar reasoning, the equation describing the response of the structure to the surrounding fluids has been deduced.

The macroscopic equations of motion of the fluid-structure system are solved here by a monolithic numerical approach. This approach consists of: a mesoscopic lattice Boltzmann method for resolving the ternary fluid flow in an Eulerian description, a finite difference method to evolve the membranes equations of motion in a Lagrangian framework, and an immersed boundary method to couple the Eulerian and Lagrangian solvers. The fluid and structure solvers are coupled through a forcing source term in the lattice Boltzmann equation, acting as a feedback of the structure's response on the flow.

We have subsequently validated our computational algorithm against Surface Evolver, an open-source software capable of modelling steady liquid surfaces problems employing an energy minimization approach. The configuration of an elastic capsule placed at a fluid-fluid interface was considered as the benchmark test. We have compared in detail the equilibrium shapes of the capsule, and its corresponding deformation parameters for different scenarios of the surface tensions and combinations of the capsule's Young's and bending moduli. An overall good agreement has been observed between our results and the reference ones. We have also demonstrated the Galilean invariance of our model equations. Finally, our algorithm has been applied to a more complex configuration, that is the capillary bridge formed between two elastic capsules. Although an extensive investigation of the parameter space was beyond the scope of the present work, it should be noted that this configuration is a particularly rich phenomenon, where the criteria for the bridge rupture and the case of unequal capsules are worth being studied in the future. 

We have assumed here that all fluid components have the same density. By modifying the Landau free-energy functional $\mathcal{E}_f$ of (\ref{eq:LandauFE}) in an appropriate way \citep{WohrwagEtAl2018} and following a rationale similar to the one presented in \S\ref{sec:Model}, our model could be extended to the case where the components of the ternary fluid mixture have different densities. The model could also be generalised so as to include more fluid components enclosed in/surrounding the elastic membranes, allowing us to tackle a wider range of applications, for example capsules containing multiple aqueous phases \citep{ScottLongEtAl2008, KusumaatmajaEtAl2009}. Also, extensions to three-dimensional configurations and other lattice Boltzmann models are straightforward. In addition, the formulations of the strain and bending energies could be readily modified to consider materials obeying different constitutive laws, such as non-linear hyperelastic materials or biological membranes. Finally, the structure solver could be adapted in order to simulate open surfaces, encountered for instance in the wetting of a soft substrate \citep{MarchandEtAl2012, StyleDufresne2012}.

\section*{Acknowledgements}
We would like to thank Professor Ken Brakke for his assistance in implementing our free energy model of the elastic membranes in Surface Evolver. MP and HK acknowledge EPSRC (EP/P007139/1) for funding. ACMS is grateful to the EPSRC Centre for Doctoral Training in Soft Matter and Functional Interfaces (EP/L015536/1) for financial support. TK acknowledges the award of a Chancellor's Fellowship from the University of Edinburgh. C. Semprebon acknowledges support from Northumbria University through the Vice-Chancellor's Fellowship Programme. 

\appendix
\section{}\label{appA}
The chemical potentials $\mu_\rho$, $\mu_\phi$ and $\mu_\psi$ are expressed as
\begin{eqnarray}\label{eq:AppA}
\mu_\rho & = & c_s^2\left(\ln\rho + 1\right) + \frac{\kappa_1}{8}\left(\rho+\phi-\psi\right)\left(\rho+\phi-\psi -1\right)\left(\rho+\phi-\psi-2\right)\nonumber\\
&& +\frac{\kappa_2}{8}\left(\rho-\phi-\psi\right)\left(\rho-\phi-\psi -1\right)\left(\rho-\phi-\psi-2\right)\nonumber\\
&& +\frac{\alpha^2}{4}\left[\left(\kappa_1+\kappa_2\right)\left(\nabla^2\psi -\nabla^2\rho\right) + \left(\kappa_2-\kappa_1\right)\nabla^2\phi \right], \\[4pt]
\mu_\phi & = & \frac{\kappa_1}{8}\left(\rho+\phi-\psi\right)\left(\rho+\phi-\psi -1\right)\left(\rho+\phi-\psi-2\right)\nonumber\\
&& -\frac{\kappa_2}{8}\left(\rho-\phi-\psi\right)\left(\rho-\phi-\psi -1\right)\left(\rho-\phi-\psi-2\right)\nonumber\\
&& +\frac{\alpha^2}{4}\left[\left(\kappa_2-\kappa_1\right)\left(\nabla^2\rho -\nabla^2\psi\right) - \left(\kappa_1+\kappa_2\right)\nabla^2\phi \right], \\[4pt]
\mu_\psi & = & -\frac{\kappa_1}{8}\left(\rho+\phi-\psi\right)\left(\rho+\phi -\psi -1\right)\left(\rho+\phi-\psi-2\right)\nonumber\\
&& -\frac{\kappa_2}{8}\left(\rho-\phi-\psi\right)\left(\rho-\phi-\psi -1\right)\left(\rho-\phi-\psi-2\right)\nonumber\\
&& +\frac{\alpha^2}{4}\left[\left(\kappa_1+\kappa_2\right)\nabla^2\rho - \left( \kappa_2-\kappa_1\right)\nabla^2\phi - \left(\kappa_1+\kappa_2+4\kappa_3 \right)\nabla^2\psi\right]\nonumber\\
&& + \kappa_3\psi\left(\psi -1\right)\left( 2\psi-1\right).
\end{eqnarray}
The pressure tensor $\mathsfbi{p}$ is given by
\begin{eqnarray}
p_{\alpha\beta} & = & p_b \delta_{\alpha\beta}+\alpha^2\kappa_{\rho\rho}\left[ \left(\partial_\alpha\rho\right)\left(\partial_\beta\rho\right)-\left(1/2\right) \left(\partial_\gamma\rho\right)^2\delta_{\alpha\beta}-\rho \left( \partial_{\gamma\gamma} \rho\right)\delta_{\alpha\beta}\right]\nonumber\\
&& +\alpha^2\kappa_{\phi\phi}\left[\left(\partial_\alpha\phi\right)\left(\partial_\beta\phi\right)-\left(1/2\right)\left(\partial_\gamma\phi\right)^2\delta_{\alpha\beta}-\phi \left( \partial_{\gamma\gamma} \phi\right) \delta_{\alpha\beta}\right]\nonumber\\
&& +\alpha^2\kappa_{\psi\psi}\left[\left(\partial_\alpha\psi\right)\left(\partial_\beta\psi\right)-\left(1/2\right)\left(\partial_\gamma\psi\right)^2\delta_{\alpha\beta}-\psi \left( \partial_{\gamma\gamma} \psi\right) \delta_{\alpha\beta}\right]\nonumber\\
&& +\alpha^2\kappa_{\rho\phi}\left[\left(\partial_\alpha\rho\right)\left(\partial_\beta\phi\right) + \left(\partial_\alpha\phi\right) \left( \partial_\beta\rho\right) - \left(\partial_\gamma\rho\right) \left( \partial_\gamma\phi\right)\delta_{\alpha\beta}\right.\nonumber\\
&& -\left.\rho\left(\partial _{\gamma\gamma}\phi\right)\delta_{\alpha\beta}-\phi\left(\partial _{\gamma\gamma}\rho\right)\delta_{\alpha\beta}\right]\nonumber\\
&& + \alpha^2\kappa_{\rho\psi}\left[\left(\partial_\alpha\rho\right)\left(\partial_\beta\psi\right) + \left(\partial_\alpha\psi\right) \left( \partial_\beta\rho\right) - \left(\partial_\gamma\rho\right) \left( \partial_\gamma\psi\right)\delta_{\alpha\beta}\right.\nonumber\\
&& -\left.\rho\left(\partial _{\gamma\gamma}\psi\right)\delta_{\alpha\beta}-\psi\left(\partial _{\gamma\gamma}\rho\right)\delta_{\alpha\beta}\right]\nonumber\\
&& + \alpha^2\kappa_{\phi\psi}\left[\left(\partial_\alpha\phi\right)\left(\partial_\beta\psi\right) + \left(\partial_\alpha\psi\right) \left( \partial_\beta\phi\right) - \left(\partial_\gamma\phi\right) \left( \partial_\gamma\psi\right)\delta_{\alpha\beta}\right.\nonumber\\
&& -\left.\phi\left(\partial _{\gamma\gamma}\psi\right)\delta_{\alpha\beta}-\psi\left(\partial _{\gamma\gamma}\right)\delta_{\alpha\beta}\right],
\end{eqnarray}
where the mixing coefficients take the form
$$\kappa_{\rho\rho} = \kappa_{\phi\phi} = \frac{\kappa_1+\kappa_2}{4},\ \kappa_{\psi\psi}= \frac{\kappa_1+\kappa_2+4\kappa_3}{4},\ \kappa_{\rho\phi} = -\kappa_{\phi\psi} = \frac{\kappa_1-\kappa_2}{4},\ \kappa_{\rho\psi} = -\frac{\kappa_1+\kappa_2}{4}.$$
The bulk pressure term $p_b$ is given by
\begin{eqnarray}
p_b & = & \rho c_s^2 + \left(\kappa_1+\kappa_2\right)\left[\frac{3}{32} \rho^4+\frac{3}{32}\phi^4+\frac{9}{16}\rho^2\phi^2+\frac{9}{16}\rho^2\psi^2 +\frac{9}{16}\phi^2\psi^2 -\frac{3}{8}\rho^3\psi\right.\nonumber\\
&& -\left.\frac{3}{8}\rho\psi^3+ \frac{3}{4}\rho^2\psi-\frac{3}{4}\rho\phi^2-\frac{3}{4}\rho\psi^2+ \frac{3}{4} \phi^2\psi-\frac{1}{4}\rho^3+\frac{1}{8}\rho^2+\frac{1}{8}\phi^2-\frac{1}{4}\rho\psi\right.\nonumber\\
&& -\left.\frac{9}{8}\rho\phi^2\psi\right]+\left(\kappa_1-\kappa_2\right)\left[ \frac{3}{8}\rho^3\phi+\frac{3}{8}\rho\phi^3-\frac{3}{8}\phi^3\psi-\frac{3}{8}\phi\psi^3-\frac{1}{4}\phi^3-\frac{3}{4}\rho^2\phi\right.\nonumber\\
&& -\left.\frac{3}{4}\phi\psi^2+\frac{1}{4}\rho\phi-\frac{1}{4}\phi\psi +\frac{9}{8}\rho\phi\psi^2-\frac{9}{8}\rho^2\phi\psi+\frac{3}{2}\rho\phi\psi\right]\nonumber\\
&& +\frac{1}{4}\left(\kappa_1+\kappa_2-8\kappa_3\right)\psi^3+\left(\kappa_1+\kappa_2+16\kappa_3\right)\left[\frac{3}{32}\psi^4+ \frac{1}{32}\psi^2\right].
\end{eqnarray}
We have ignored terms that provide constant pressure contribution throughout the whole system.

\bibliographystyle{jfm}
\bibliography{Article}

\end{document}